\documentclass[reqno]{amsart}
\usepackage{amsmath}
\usepackage{amsthm}
\usepackage{amssymb}
\usepackage[left=1.0in, right=1.0in]{geometry}
\usepackage{tikz}
\usepackage[all]{xy}
\usepackage{graphicx}
\usepackage{mathrsfs}
\usepackage{hyperref}
\usepackage{microtype}
\usepackage{url,doi}
\usepackage{braket}
\usepackage{breakurl}
\usepackage{ifpdf}
\usepackage{mathtools}
\usepackage{color,soul}
\usepackage{tikz}
\usepackage[all]{xy}
\usepackage{graphicx}
\usepackage{braket}
\usepackage{ifpdf}
\usepackage{mathtools}
\usepackage{color}
\usepackage{rotating}
\newcommand{\ba}{\begin{align}}
\newcommand{\ea}{\end{align}}
\newcommand{\be}{\begin{equation}}
\newcommand{\ee}{\end{equation}}
\newcommand{\blue}[1]{{\color{blue}#1}}
\newcommand{\beq}{\begin{eqnarray}}
\newcommand{\eeq}{\end{eqnarray}}

\newcommand{\beqs}{\begin{eqnarray*}}
\newcommand{\eeqs}{\end{eqnarray*}}

\newcommand\FS{\mathfrak{F}_S}
\newcommand\Z{\mathbb{Z}}
\tikzstyle WL=[line width=3pt,opacity=1.0]
\newcommand{\drawWL}[3]
{
    \draw[white,WL]  (#2) -- (#3);
    \draw[#1] (#2) -- (#3);
}

\newcommand{\Max}{\ket{\rm{Max}}}
\newcommand{\GHZ}{\ket{\rm{GHZ}}}

\theoremstyle{plain}

\begin{document}
\pagenumbering{gobble}
\title{Quon 3D Language for Quantum Information}
\author[Zhengwei Liu, Alex Wozniakowski, and Arthur Jaffe]{Zhengwei Liu, Alex Wozniakowski, and Arthur Jaffe\\Harvard University}

\address{
\href{mailto:zhengweiliu@fas.harvard.edu}{ZhengweiLiu@fas.harvard.edu}}
\address{
\href{mailto:airwozz@gmail.com}{airwozz@gmail.com}}
\address{\href{mailto:arthur_jaffe@harvard.edu}{Arthur\_Jaffe@harvard.edu}}

\keywords{Quon Language  $|$ Picture-Language $|$ Quantum Information $|$ Joint Relation $|$ Topological Isotopy $|$ Topological Algebra \quad \textit{Correspondence}: \href{mailto:arthur_jaffe@harvard.edu}{Arthur\_Jaffe@harvard.edu}}

\begin{abstract}
We present a  3D, topological picture-language for quantum information.  Our approach combines charged excitations carried by strings, with topological properties that arise from embedding the strings in the interior of a three-dimensional manifold with boundary. A quon is a composite that acts as a particle. Specifically a quon is a hemisphere containing a neutral pair of open strings with opposite charge.  We interpret multi-quons and their transformations in a natural way.  We obtain a new type of relation, a string-genus ``joint relation,'' involving both a string and the 3D manifold.
We use the joint relation to obtain a topological interpretation of the $C^{*}$ Hopf algebra relations, that are widely used in tensor networks.  We obtain a 3D representation of the Controlled NOT or CNOT gate (that is considerably simpler than earlier work) and a 3D topological protocol for teleportation.
\end{abstract}

\maketitle
\section*{Significance}
\noindent We give a new three-dimensional, picture-language for quantum information.  This language is based on an inherently 3D pictorial representation of particle-like excitations (quons), and of transformations acting on them. Mathematical identities and quantum information protocols are expressed through deformations of these pictures.  We explore our language, highlighting conceptual insights, 3D visualizations, and suggestive intuition that it motivates for understanding algebra and quantum information.

\section{Introduction}
Topological quantum information was formulated by Kitaev \cite{Kitaev03} and Freedman et al. \cite{Freedman-etal}.  Here we formulate a 3D topological picture-language that we call the \textit{quon language}---suggesting  quantum particles. It  leads to strikingly elementary mathematical proofs and insights into quantum information protocols. 
In our previous work we represented qudits, the basic unit of quantum information, using charged strings in 2D. This fits naturally into the framework of planar para algebras~\cite{JL,JLW-QI, JLW-HS,JLW-TDP}.  We call this our \textit{two-string model}.  

We also found a  \textit{four-string model} in 2D, in which we represent a $1$-qudit vector as a neutral pair of  particle-antiparticle charged strings~\cite{JL,JLW-QI}. These charged strings have the properties of parafermions.  The presence of charges lead to para-isotopy relations, which reflect the parafermion multiplication laws.  Neutral pairs satisfy isotopy, a very appealing property.  However, braiding two strings from different qudits destroys individual qudit neutrality, and this problem seemed unsurmountable for multi-qudit states.  So can one isolate those transformations that map the neutral pairs into themselves?

Here we solve this problem by defining \textit{quons} and the quon picture-language.  We embed the neutral pairs of charged strings representing qudits, into the interior of a $3$-manifold.  The quon language has the flavor of a topological field theory with strings.  The resulting composites of 3-manifolds and strings give us quon states, transformations of quons, and quon measurements. However the composites contain a further new aspect: there are topological relations that involve both the strings and the manifolds.  We call them \textit{joint relations}.  These joint relations provide basic grammatical structure as well as insight into our language.

In \S\ref{Sect:String-Genus}, we see that if a neutral string surrounds a genus in the manifold, then one can remove them both.    In \S\ref{Sect:TopHopf} we use this joint relation to obtain an elementary understanding of Frobenius and $C^{*}$-Hopf algebra relations stated in \S\ref{Sec:Bi Frobenius algebra}. These relations are key in tensor network theory. In \S\ref{Sect:UMTC} we see that our construction even provides four-string structure for unitary modular tensor categories, with quons a special case.

\section{Basic Grammar}
\pagenumbering{arabic}
\subsection{The ${1}$-quon space}
\setcounter{page}{2}
We represent a  $1$-quon by a heimisphere, with no input points and  four output points.  Transformations of $1$-qudits have four input and four output points in a cylinder, so we call this a \textit{four-string model}.  
We represent a 1-quon measurement by a hemisphere with four input points and no output points.

In case a quon is a qudit of degree $d$, one has a simple representation for a $1$-quon basis: the interior of a hemisphere contains two charged strings, each linking two of the output points.  The value of the charge on one sting may equal either $0,1,\ldots,d-1\in \mathbb{Z}_{d}$, while the other string carries the negative of that charge.  For $d=2$ the quons reduce to Majorana fermions.

The four-string model for a qudit found in \S5.3 of~\cite{JL} arose as a  natural generalization of Kitaev's picture~\cite{Kitaev} of a spin as a pair of fermions. The four strings arise as we represent the Pauli matrices $X,Y,Z$ by four parafermions.
In our reinterpretation, we replace the two fermions by a pair of parafermion/anti-parafermion unitaries with opposite charge.
We represent transformations on $1$-quons as a box with four input points and four output points, embedded in a three-manifold.  We describe various bases in \S\ref{Sec:1QuonBasis}.

\subsection{Multi-quon space}
Multi-quons have a hemisphere for each $1$-quon.  A transformation on $n$-quons has charged strings in a three ball with $n$ input handles and $n$ output handles, each containing four strings. This representation leads to a natural multi-particle structure; it allows us to analyze the full Hilbert space for multi-quons, with each individual quon remaining a neutral pair.
\subsection{Quons as topological algebra}\label{Sec:TopAlg}
Picture-language for tensor networks arose in Penrose~\cite{Penrose}, Deutsch~\cite{Deutsch}, and in D\"ur, Vidal, and Cirac~\cite{DVC}.  The Hopf algebra axioms were studied in tensor networks by Lafont~\cite{Lafont}.   Abramsky, Coecke, and others studied quantum information extensively from a categorical point of view,  and found many applications in tensor networks~\cite{AbramskyCoecke04,Coecke-Duncan-1,Coecke-Duncan-2,Coeckebook,BB,DBJC,Backens}.  
Vicary and Reutter applied 2-categories and biunitaries in planar algebras to quantum information~\cite{Vicary,Vicary-Reutter}.

%

%

Our quons live in  3D space, and thereby capture categorical structures in two directions.  We obtain Frobenius algebras in the $X$-direction and the $Y$-direction, corresponding respectively to the COPY and SUM maps in tensor networks. 
We explain these concepts in~\S\ref{Sec:Bi Frobenius algebra}.
They define the underlying Hilbert space as a $C^*$-Hopf algebra, as explained in~\S\ref{Sect:TopHopf}. Moreover, the string Fourier transform $\FS$ is a $90^\circ$ rotation around the $Z$-axis. Conjugation by $\FS$ maps one Frobenius algebra to the other. This gives a topological interpretation of the algebraic axioms of Hopf algebras.

Algebraic relations arise from the  invariance of certain elementary diagrams under topological isotopy.
It is significant that different algebraic conditions have the same topological representation.  In other words, diagrams that are equivalent up to isotopy can have different algebraic meanings
when they are located at different positions.
We have already used this philosophy in our two-string model
 to design protocols in a topological way  \cite{JLW-HS,JLW-TDP}.  Although  3D braiding appears in many places, e.g. \cite{Kitaev03,Jones,Freedman-etal,KL-1,RHG,HFDV,BLKW}, we believe that our present work is the first to combine charged strings with 3D manifolds.

Using these diagrams, in \S\ref{Sect:Teleportation} we obtain a 3D representation of the CNOT map and the quantum teleportation protocol. The teleportation protocol becomes a topological protocol in the quon language.

\section{Parafermion algebras}
The \textit{parafermion algebra} $PF_n$ of order $d$ is a $\ast$-algebra with unitary generators $c_{m}$, $m=1,2,\cdots,n$, which satisfy
\begin{equation}\label{ParafermionAlgebra}
c_{m}^{d}=1 \; \; \; \mbox{and} \; \; \; c_{m}c_{m'}=q \, c_{m'}c_{m} \; \; \; \mbox{for} \; \; 1 \leq  m< m' \leq n.
\end{equation}
Here $q \equiv e^{\frac{2 \pi i}{d}}$ and $i \equiv \sqrt{-1}$.
Consequently $c^{\ast}_{j}=c^{-1}_{j}=c^{d-1}_{j}$, where $\mbox{*}$ denotes the adjoint.
Majorana fermions arise from the case $d=2$.
The Jordan-Wigner transformation gives the isomorphism  $PF_{2n} \cong M_d(\mathbb{C})^{\otimes n}$ to tensor products of $d\times d$ matrices.
The parafermion algebra $PF_n$ has a basis $\displaystyle \{ c^{\alpha}\}$.
Here
$
c^{\alpha} = c_{1}^{\alpha_{1}}\cdots c_{n}^{\alpha_{n}}$,
and $\alpha_k \in \Z_d$.
The charge of $c^{\alpha}$ is defined to be $|c^\alpha|=\sum_{k=1}^{n} \alpha_k$ in $Z_d$.
The zero-charged elements $PF^{0}_{n}$ form a subalgebra of the parafermion algebra $PF_{n}$, namely the neutral subalgebra.

\subsection{Parafermion planar para algebras (PAPPA)}
Given $1\leqslant m\leqslant n$, our diagrammatic representation for $c_{m}^{\alpha_{m}}$, with $\alpha_{m}=k$, as explained in \cite{JL}, is
$$
c_m^k \longleftrightarrow
\raisebox{-.3cm}
{\scalebox{.7}{
\begin{tikzpicture}
\node at (-.75,.5) {$...$};
\node at (.75,.5) {$...$};
 \draw (-1,0)--(-1,1);
 \draw (1,0)--(1,1);
 \draw (-.5,0)--(-.5,1);
 \draw (.5,0)--(.5,1);
 \draw (0,0)--(0,1);
  \node at (-.2,.5) {k};
\end{tikzpicture}}}\;,
$$
where the $m^{\rm th}$ string is labeled by $k$.
In our notation we have the relations
\beq\label{AddCharge}
\textbf{Multiplication:  }&&
\raisebox{-.25cm}
{\scalebox{.7}{
\begin{tikzpicture}
\draw (0,0) --(0,1);
\node (0,0) at (-0.2,0.8) {$\ell$};
\node (0,0) at (-0.2,0.2) {$k$};
\node (0.45,0.25) at (0.45,0.5) {$=$};
\node (0.9,0.35) at (1.4,0.5) {$k + \ell$};
\draw (2,0) --(2,1);
\node (1.4,-0.05) at (2.3,0.3) {,};
\end{tikzpicture} }}\qquad
\raisebox{-.25cm}
{\scalebox{.7}{
\tikz{
\node (3.7,0.35) at (4.8,0.5) {$d$};
\draw (5,0) --(5,1);
\node (4.4,0.25) at (5.4,0.5) {$=$};
\draw (5.8,0) --(5.8,1);
\node (4.4,0.25) at (6.2,0.5) {$=$};
\node (3.7,0.35) at (6.5,0.5) {$0$};
\draw (6.7,0) --(6.7,1);
\node (4.95,0) at (7,0.3) {.};
}}}\nonumber\\
%
\textbf{Para isotopy:  }&&
\raisebox{-.25cm}
{\scalebox{.7}{
\begin{tikzpicture}
\draw (0,0) --(0,01);
\draw (0.2,0) --(0.2,01);
\node at (.5,.5) {$\cdots$};
\node (0,0) at (-0.15,0.2) {$k$};
\draw (0.8,0) --(0.8,1);
\draw (1.15,0) --(1.15,1);
\node (1.2,0) at (1,0.8) {$\ell$};
\end{tikzpicture}
}}
=  \ \scriptstyle q^{k\ell}
\raisebox{-.25cm}
{\scalebox{.7}{
\begin{tikzpicture}
\draw (0,0) --(0,01);
\draw (0.2,0) --(0.2,01);
\node at (.5,.5) {$\cdots$};
\node (0,0) at (-0.15,0.8) {$k$};
\draw (0.8,0) --(0.8,1);
\draw (1.15,0) --(1.15,1);
\node (1.2,0) at (1,0.2) {$\ell$};
\end{tikzpicture}}}\;.
\eeq
The strings between $k$-charged and $\ell$-charged strings are not charged. 
Take $\zeta$ to be a square root of $q$, such that
	\be
	\zeta^{d^2}=1\;.
	\ee
We can  interpolate between the diagrams in \eqref{AddCharge} as a
\be \label{Equ:twisted tensor product}
\hskip -1.5cm
\textbf{Twisted product:    }\qquad
\raisebox{-.25cm}
{\scalebox{.7}{
\begin{tikzpicture}
\draw (0,0) --(0,01);
\draw (0.2,0) --(0.2,01);
\node at (.5,.5) {$\cdots$};
\node (0,0) at (-0.15,0.5) {$k$};
\draw (0.8,0) --(0.8,1);
\draw (1.15,0) --(1.15,1);
\node (1.2,0) at (1,0.5) {$\ell$};
\end{tikzpicture}
}}
:=  \ \scriptstyle \zeta^{k\ell}
\raisebox{-.25cm}
{\scalebox{.7}{
\begin{tikzpicture}
\draw (0,0) --(0,01);
\draw (0.2,0) --(0.2,01);
\node at (.5,.5) {$\cdots$};
\node (0,0) at (-0.15,0.8) {$k$};
\draw (0.8,0) --(0.8,1);
\draw (1.15,0) --(1.15,1);
\node (1.2,0) at (1,0.2) {$\ell$};
\end{tikzpicture}}}\;.
\ee
In the PAPPA model, the charged strings satisfy:
\begin{itemize}
\setlength\itemsep{1em}

\item
\textcolor{black}{$\rho_{\pi}(c^j)=\zeta^{j^2} c^j,$}
\noindent where $\rho_{\pi}$ is a rotation by $\pi$ on the plane. Then $\rho_{2\pi}(c^j)=q^{j^2} c^j$ for the $2\pi$ rotation $\rho_{2\pi}$.

\item \raisebox{-.3cm}{
\begin{tikzpicture}
  \draw (0,0) circle (.4);
  \node at (-.6,0) {j};
\end{tikzpicture}
}
$=0,~1\leq j\leq d-1$, and
%
%
%
\raisebox{-.3cm}{
\begin{tikzpicture}
  \draw (0,0) circle (.4);
\end{tikzpicture}
}
$=\delta=\sqrt{d}.$

\item{}
Let $\omega=\frac{1}{\sqrt{d}}\sum_{j=0}^{d-1}\zeta^{j^{2}}$.  This  is a phase, as shown in Proposition 2.15 of \cite{JL}.  Then
\beq
\raisebox{-.4cm}{
\begin{tikzpicture}
  \draw (0,0)--(0,1);
  \draw (.6,0)--(.6,1);
\end{tikzpicture}}
&=&\frac{1}{\sqrt{d}}\sum_{j=0}^{d-1}
\raisebox{-.5cm}{
\begin{tikzpicture}
  \draw (0,0)--++(0,.1) arc (180:0:.3)--++(0,-.1);
  \draw (0,1)--++(0,-.1) arc (-180:0:.3)--++(0,.1);
  \node at (.5-.2,0.1) {j};
  \node at (.5-.2,0.9) {-j};
\end{tikzpicture}}\nonumber\\
\displaystyle \raisebox{-.4cm}{
\begin{tikzpicture}
  \draw (0,0)--(.6,1);
  \drawWL {}{.6,0}{0,1};
\end{tikzpicture}}
&=&\frac{1}{\sqrt{\omega d}}
\sum_{j=0}^{d-1} \zeta^{j^2}
\raisebox{-.5cm}{
\begin{tikzpicture}
  \draw (0,0)--++(0,.1) arc (180:0:.3)--++(0,-.1);
  \draw (0,1)--++(0,-.1) arc (-180:0:.3)--++(0,.1);
  \node at (.5-.2,0.1) {j};
  \node at (.5-.2,0.9) {-j};
\end{tikzpicture}}
\ .\nonumber
\eeq
\end{itemize}
Using the above definition of braiding, we can establish the braided relation that any neutral diagram can move above or under the strings, see \cite{JLW-HS}.
Therefore the neutral diagrams can be lifted to the 3D space.

\subsection{Categorical approach to the neutral part of the parafermion algebra}
For  readers who are familiar with category theory, one can consider the neutral diagrams as morphisms in a monoidal category.
The neutral part of PAPPA is the $\Z_{d}$ unshaded subfactor planar algebra. It is a $Z_{2}$ graded unitary fusion category. Its even part is the monoidal category $Vec_{Z_{d}}$, whose simple objects $X_g$ are labeled by group elements $g$ in $\Z_{d}$ indicating the fusion rule.
It has only one odd simple object $\tau=\overline{\tau}$, where $\overline{\tau}$ is the dual of $\tau$, such that $\tau^{2}=\gamma=\bigoplus_{g \in Z_{d}} X_{g}$. Thus $\gamma$ is a Frobenius algebra.
Then the neutral subalgebra of the parafermion algebra is given by $PF^{0}_{n}=\hom(\tau^{n},\tau^{n})$.

\section{Details of the quon model}
%
\subsection{Quons}
An $n$-quon is represented by $n$ hemispheres.  We call the flat disc on the boundary of each hemisphere, a boundary disc. Each hemisphere contains a neutral diagram with four boundary points  on its boundary disc.
The dotted box designates the internal structure  that specifies the quon vector.
For example, the 3-quon is represented as
\vskip -.3cm
\be \label{Equ:qudit}
\scalebox{.5}
{
\raisebox{-1.cm}{
\begin{tikzpicture}
\begin{scope}[xscale=.5,yscale=.25]
\foreach \x in {0,6,12}{
\draw[blue] (\x,12) arc (-180:0:2.5);
\draw[blue] (\x,12) arc (180:0:2.5 and 6);
\draw[blue,dashed] (\x,12) arc (180:0:2.5);
\draw[dashed] (\x+.75,13) rectangle (\x+4.25,15);
\foreach \y in {1,2,3,4}
{
\draw (\x+\y,12)--++(0,1);
}
}
\node at (0+2.5,14) {$v_{1}$};
\node at (6+2.5,14) {$v_{2}$};
\node at (12+2.5,14) {$v_{3}$};
\end{scope}
\end{tikzpicture}}}\ .
\ee
Here $v_{j}$ labels a $1$-qudit vector given by neutral diagrams with four boundary points in the hemisphere. Here we orient the boundary disc of the hemispheres to lie on the $X$-$Y$-plane in the 3D space.

\subsection{Transformations}
An $n$-quon transformation is represented by a neutral element $T$ in $PF_{4n}=\hom(\tau^{4n},\tau^{4n})$ embedded in a 3-manifold, isotopic to a 3D ball.
The 3-manifold has $n$ boundary discs on the top and $n$ at the bottom. Each disc contains four boundary points of $T$.
For example,
a $3$-quon transformation $T$ will have the representation:
\vskip -.4cm
\be \label{Equ:T}
\scalebox{.65}{$
\raisebox{-2cm}{
\begin{tikzpicture}
\begin{scope}[xscale=.5,yscale=.25]
\draw[dashed] (.75,4) rectangle (16.25,8);
\node at (8.5,6) {$T$};
\foreach \x in {1,2,3,4,7,8,9,10,13,14,15,16} {
\draw (\x,0)--++(0,4);
\draw (\x,8)--++(0,4);
}

\foreach \x in {0.5,6.5,12.5}{
\draw[white,WL]  (\x,12) arc (-180:180:2);
\draw[blue] (\x,12) arc (-180:180:2);
\draw[blue] (\x,0) arc (-180:0:2);
\draw[blue,dashed] (\x,0) arc (180:0:2);
}
\draw[blue] (0.5,0)--++(0,12);
\draw[blue] (16.5,0)--++(0,12);
\draw[blue] (4.5,12)--++(0,-2) arc (-180:0:1)--++(0,2);
\draw[blue] (10.5,12)--++(0,-2) arc (-180:0:1)--++(0,2);
\draw[blue] (4.5,0)--++(0,2) arc (180:0:1)--++(0,-2);
\draw[blue] (10.5,0)--++(0,2) arc (180:0:1)--++(0,-2);
\end{scope}
\end{tikzpicture}}$}\ .
\ee

\subsection{Isotopy of Neutral Diagrams}
In addition to the relations for charged strings, we allow isotopy of strings in 3-manifolds. We define a relation for 3-manifolds: if a 3D-ball has no diagram inside, then it can be removed.
Moreover we define a joint relation between diagrams and 3-manifolds.

Suppose $T\in \hom(\tau^m,\tau^n)$. Let $B_m$ be an orthonormal basis (ONB) of $\hom (\tau^m,\tau^m)$, and let $B_n$ be an ONB of $\hom (\tau^n,\tau^n)$.
We define the relation
\be\label{Equ:joint}
\scalebox{.65}{$
\raisebox{-2cm}{
\begin{tikzpicture}
\begin{scope}[xscale=.5,yscale=.25]
\draw[dashed] (.75,4) rectangle (4.25,8);
\node at (2.5,6) {$T$};
\foreach \x in {1,4} {
\draw (\x,0)--++(0,4);
\draw (\x,8)--++(0,4);
}
\node at (2.5,0) {$\cdots$};
\node at (2.5,12) {$\cdots$};
\foreach \x in {0.5}{
\draw[white,WL]  (\x,12) arc (-180:180:2);
\draw[blue] (\x,12) arc (-180:180:2);
\draw[blue] (\x,0) arc (-180:0:2);
\draw[blue,dashed] (\x,0) arc (180:0:2);
}
\draw[blue] (.5,0)--(.5,12);
\draw[blue] (4.5,0)--(4.5,12);
\end{scope}
\end{tikzpicture}
}
:= {\displaystyle \sum_{\alpha\in B_m}}
\raisebox{-2cm}{
\begin{tikzpicture}
\begin{scope}[xscale=.5,yscale=.25]
\draw[dashed] (.75,1) rectangle (4.25,3);
\draw[dashed] (.75,4) rectangle (4.25,6);
\draw[dashed] (.75,8) rectangle (4.25,10);
\node at (2.5,2) {$T$};
\node at (2.5,5) {$\alpha$};
\node at (2.5,9) {$\alpha^*$};
\foreach \x in {1,4} {
\draw (\x,0)--++(0,1);
\draw (\x,3)--++(0,1);
\draw (\x,10)--++(0,2);
}
\node at (2.5,0) {$\cdots$};
\node at (2.5,12) {$\cdots$};
\foreach \x in {0.5}{
\draw[white,WL]  (\x,12) arc (-180:180:2);
\draw[blue] (\x,12) arc (-180:180:2);
\draw[blue] (\x,0) arc (-180:0:2);
\draw[blue,dashed] (\x,0) arc (180:0:2);
}
\draw[blue] (.5,0)--++(0,6) arc (180:0:2 and .5)--++(0,-6);
\draw[blue] (.5,12)--++(0,-4) arc (-180:0:2 and .5)--++(0,4);
\end{scope}
\end{tikzpicture}
}
=
{\displaystyle \sum_{\beta\in B_n}}
\raisebox{-2cm}{
\begin{tikzpicture}
\begin{scope}[xscale=.5,yscale=.25]
\draw[dashed] (.75,1) rectangle (4.25,3);
\draw[dashed] (.75,5) rectangle (4.25,7);
\draw[dashed] (.75,8) rectangle (4.25,10);
\node at (2.5,9) {$T$};
\node at (2.5,2) {$\beta$};
\node at (2.5,6) {$\beta^*$};
\foreach \x in {1,4} {
\draw (\x,0)--++(0,1);
\draw (\x,7)--++(0,1);
\draw (\x,10)--++(0,2);
}
\node at (2.5,0) {$\cdots$};
\node at (2.5,12) {$\cdots$};
\foreach \x in {0.5}{
\draw[white,WL]  (\x,12) arc (-180:180:2);
\draw[blue] (\x,12) arc (-180:180:2);
\draw[blue] (\x,0) arc (-180:0:2);
\draw[blue,dashed] (\x,0) arc (180:0:2);
}
\draw[blue] (.5,0)--++(0,3) arc (180:0:2 and .5)--++(0,-3);
\draw[blue] (.5,12)--++(0,-7) arc (-180:0:2 and .5)--++(0,7);
\end{scope}
\end{tikzpicture}
}$}
\ee

Basic linear algebra shows that the second equality always holds.
It says that the relation is well-defined up to isotopy of neutral diagrams in 3-manifolds.
By this relation, the picture \eqref{Equ:T} reduces to a linear sum of pictures of the following form:
\be \label{Equ:transformation-basis}
\raisebox{-1.3cm}{
\scalebox{.65}{$
\begin{tikzpicture}
\begin{scope}[shift={(0,10)},xscale=.5,yscale=-.25]
\foreach \x in {0,6,12}{
\draw[dashed] (\x+.75,13) rectangle (\x+4.25,15);
\foreach \y in {1,2,3,4}
{
\draw (\x+\y,12)--++(0,1);
}
\draw[blue] (\x,12) arc (-180:0:2.5);
\draw[blue] (\x,12) arc (180:0:2.5 and 6);
\draw[blue] (\x,12) arc (180:0:2.5);
}
\node at (0+2.5,14) {$w^*_{1}$};
\node at (6+2.5,14) {$w^*_{2}$};
\node at (12+2.5,14) {$w^*_{3}$};
\end{scope}
\begin{scope}[xscale=.5,yscale=.25]
\foreach \x in {0,6,12}{
\draw[blue] (\x,12) arc (-180:0:2.5);
\draw[blue] (\x,12) arc (180:0:2.5 and 6);
\draw[blue,dashed] (\x,12) arc (180:0:2.5);
\draw[dashed] (\x+.75,13) rectangle (\x+4.25,15);
\foreach \y in {1,2,3,4}
{
\draw (\x+\y,12)--++(0,1);
}
}
\node at (0+2.5,14) {$v_{1}$};
\node at (6+2.5,14) {$v_{2}$};
\node at (12+2.5,14) {$v_{3}$};
\end{scope}
\end{tikzpicture}$}}
\ee

If we take $v_i,w_i$ to be elements in an ONB of $\hom(1,\tau^4)$, then these pictures represent matrix units of qudit transformations.
Therefore we obtain a representation of quons and transformations by neutral diagrams in 3-manifolds modulo relations.
To simplify the notations, sometimes we ignore the 3-manifold, if there is no confusion.

\section{${1}$-Quon bases}\label{Sec:1QuonBasis}
\subsection{Qubit case}
The space of $1$-qubit states is known as the Bloch sphere. Vector states lie on the surface. The antipodes for a unit $3$-vector $\pm\vec n$ are assigned the eigenvectors of $n_{x}X + n_{y}Y+n_{z}Z$, where $X,Y,Z$ are the Pauli matrices. The eigenvalues are $\pm1$. The usual convention is to let the eigenvectors of $Z$ be  $\ket{0}=\ket{0_{Z}}$ at the south pole and $\ket{1}=\ket{1_{Z}} $ at the north pole. Then there are three fundamental sets of bases of the $1$-qubit space for $n_{x}=1$, etc.  They are  $\ket{0_X}=\frac{1}{\sqrt{2}}(\ket{0}+\ket{1})$,  $\ket{1_X}=\frac{1}{\sqrt{2}}(\ket{0}-\ket{1})$, and likewise $  \ket{0_Y}=\frac{1}{\sqrt{2}}(\ket{0}+i\ket{1})$, $\ket{1_Y}=\frac{1}{\sqrt{2}}(\ket{0}-i\ket{1})$.  The Bloch sphere can be drawn  as:
%
%
$$\begin{tikzpicture}
\raisebox{0cm}{
  \draw (-1,0) arc (180:-180:1);
  \draw (-1,0) to [bend right=30] (1,0);
  \draw [dashed] (-1,0) to [bend right=-30] (1,0);
  \draw (0,1) to [bend right=30] (0,-1);
  \draw [dashed] (0,1) to [bend right=-30] (0,-1);
  \node at (1.8,0) {$\ket{0}+\ket{1}$};
  \node at (-1.8,0) {$\ket{0}-\ket{1}$};
  \node at (0,1.5) {$\ket{1}$};
  \node at (0,-1.5) {$\ket{0}$};
  \node at (1.2,1) {$\ket{0}+i\ket{1}$};
  \node at (-1.2,-1) {$\ket{0}-i\ket{1}$};
}
\end{tikzpicture}$$

\subsection{The general ${1}$-quon case}\label{Sect:1-QuonBasis}
In the quon model, the three different ways of connecting the four boundary points give the $X$, $Y$, $Z$-basis of the 1-quon space.  For $k=0,1,\ldots,d-1\in \mathbb{Z}_{d}$,
\begin{itemize}
\item Z-basis:
\raisebox{-.27cm}{$
\begin{tikzpicture}
\begin{scope}[shift={(4,-.5)},xscale=.4,yscale=.4]
\node at (-1.5,.5-1) {$\ket{k_Z}=\frac{1}{\sqrt{d}}\ \ $};
\draw (1,-1)--++(0,1) arc (180:0:.5)--++(0,-1);
\draw (3,-1)--++(0,1) arc (180:0:.5)--++(0,-1);
\node at (.5,-.5) {k};
\node at (3.5,-.5) {-k};
\end{scope}
\end{tikzpicture}
$}

\item X-basis:
\raisebox{-.56cm}{$
\begin{tikzpicture}
\begin{scope}[shift={(4,-.5)},xscale=.4,yscale=.4]
\node at (-1.5,.5) {$\ket{k_X}=\frac{1}{\sqrt{d}}\ \ $};
\draw (1,-1)--++(0,1) arc (180:0:1.5)--++(0,-1);
\draw (2,-1)--++(0,1) arc (180:0:.5)--++(0,-1);
\node at (.5,-.5) {k};
\node at (1.5,-.5) {-k};
\end{scope}
\end{tikzpicture}
$}

\item Y-basis:
\raisebox{-.39cm}{$
\begin{tikzpicture}
\begin{scope}[shift={(4,-.5)},xscale=.4,yscale=.4]
\node at (-1.5,.5-.5) {$\ket{k_Y}=\frac{1}{\sqrt{d}}\ \ $};
\draw (1,-1)--++(0,1) arc (180:0:1)--++(0,-1);
\draw[white,WL] (2,-1)--++(0,1) arc (180:0:1)--++(0,-1);
\draw (2,-1)--++(0,1) arc (180:0:1)--++(0,-1);
\node at (.5,-.5) {k};
\node at (1.5,-.5) {-k};
\end{scope}
\end{tikzpicture}
$}

\end{itemize}
These are the three eigenbases of the three unitary Pauli matrices.  The matrices are given diagrammatically  in \cite{JL}, respectively as
\be\label{Eq:Pauli}
{Z}=
\raisebox{-.4cm}{
\tikz{
\draw (-2/3,0)--(-2/3,1);
\draw (-3/3,0)--(-3/3,1);
\draw (-4/3,0)--(-4/3,1);
\draw (-5/3,0)--(-5/3,1);
\node at (-9/6,1/2) {\small{-1}};
\node at (-11/6,1/2) {\small{1}};
}} \;,\quad
{X}=
\raisebox{-.4cm}{
\tikz{
\draw (-2/3,0)--(-2/3,1);
\draw (-3/3,0)--(-3/3,1);
\draw (-4/3,0)--(-4/3,1);
\draw (-5/3,0)--(-5/3,1);
\node at (-5/6,1/2) {\small{-1}};
\node at (-11/6,1/2) {\small{1}};
}} \;, \quad
{Y}=
\raisebox{-.4cm}{
\tikz{
\draw (-2/3,0)--(-2/3,1);
\draw (-3/3,0)--(-3/3,1);
\draw (-4/3,0)--(-4/3,1);
\draw (-5/3,0)--(-5/3,1);
\node at (-7/6,1/2) {\small{1}};
\node at (-11/6,1/2) {\small{-1}};
}} \;.
\ee

In the quon model we represent the basis of the 1-quon space by a pair of strings with opposite charges,  embedded in a hemisphere, and exiting the bottom.
The algebraic adjoint operation is given by a charge-inverting, geometric reflection along the $Z$-direction.
Therefore a measurement is represented by a pair of strings with opposite charges in the reflected hemisphere:
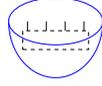
\begin{center}
\scalebox{.5}{
\begin{tikzpicture}
\begin{scope}[xscale=.5,yscale=-.25]
\foreach \x in {0}{
\draw[blue] (\x,12) arc (-180:0:2.5);
\draw[blue] (\x,12) arc (180:0:2.5 and 6);
\draw[blue] (\x,12) arc (180:0:2.5);
\draw[dashed] (\x+.75,13) rectangle (\x+4.25,15);
\foreach \y in {1,2,3,4}
{
\draw (\x+\y,12)--++(0,1);
}
}
\node at (0+2.5,14) {};
\end{scope}
\end{tikzpicture}}\ .
\end{center}
The charge represents the result of the measurement.

\section{$1$-quon Clifford group}
The 1-quon transformations $\{X,Y,Z,F,G\}$ are generators of the 1-quon Clifford group. Their algebraic definitions are given by:
	\beq
		X\ket{k} &=& \ket{k+1}\;, \
		Y\ket{k} = \zeta^{1-2k} \ket{k-1}\;,\
		Z\ket{k} = q^{k} \ket{k}\;, \nonumber\\
        F\ket{k} &=&\frac{1}{\sqrt{d}}\sum_{l=0}^{d-1} q^{kl}\ket{l} \;, \qquad
        G\ket{k} =\zeta^{k^2}\ket{k} \;.
    \eeq
When we consider a 1-quon as a vector state, the quon transformations are defined up to a phase. These 1-quon transformations form a group $\Z_d^2 \rtimes SL(2,\Z_d)$ as shown in \cite{JL}.  The Pauli matrices are given diagrammatically by \eqref{Eq:Pauli}, while
%
%
\be
F=~\raisebox{-.4cm}{\tikz{
\begin{scope}[xscale=1,yscale=1]
\draw (-3/3,0)--(-2/3,1);
\draw (-4/3,0)--(-3/3,1);
\draw (-5/3,0)--(-4/3,1);
\drawWL {}{-2/3,0}{-5/3,1};
\end{scope}
}}
=~\raisebox{-.4cm}{\tikz{
\draw (-2/3,0)--(-5/3,1);
\drawWL {}{-3/3,0}{-2/3,1};
\drawWL {}{-4/3,0}{-3/3,1};
\drawWL {}{-5/3,0}{-4/3,1};
}}
\;,
\text{~}
G=\raisebox{-.4cm}{
\tikz{
\draw (-2/3,0)--(-2/3,1);
\draw (-3/3,0)--(-3/3,1);
\draw (-5/3,0)--(-4/3,1);
\drawWL {}{-4/3,0}{-5/3,1};
}}=
\raisebox{-.4cm}{
\tikz{
\draw (-4/3,0)--(-4/3,1);
\draw (-5/3,0)--(-5/3,1);
\draw (-3/3,0)--(-2/3,1);
\drawWL {}{-2/3,0}{-3/3,1};
}} \;.
\ee

\section{$n$-Quon Clifford group}
Let $C_X$ be the controlled $X$ transformation. For the qubit case it becomes CNOT.
The $n$-qudit Clifford group is generated by $\{X,Y,Z,F,G,C_X\}$.
We represent the qudit transformation $C_X$ by neutral diagrams in 3-manifolds.

It is more natural to represent these neutral diagrams in the 3D space. In the 3D space, we label the four boundary points as 1,2,3,4 corresponding to the order of the boundary points of the 2D diagrams.
The order indicates the choice of basis in the 3D space.
We discuss more about $C_X$ and the 3D representations in \S \ref{Sec:Bi Frobenius algebra}--\ref{Sect:TopHopf}.
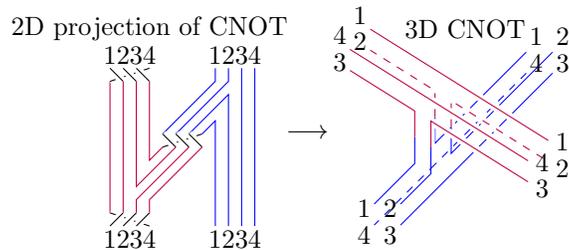
\begin{center}
\begin{tikzpicture}
\begin{scope}[xscale=.7,yscale=.7]
\node at (6,0) {$~$};
\node at (-2,4) {2D projection of CNOT};
\begin{scope}[shift={(0,0)},xscale=-.5,yscale=.5]
\begin{scope}[shift={(0,0)},xscale=.5,yscale=.5]
\draw[blue] (0,1)--(0,13);
\draw[blue] (1,1)--(1,13);
\draw[blue] (2,1)--(2,11)--(5,8);
\draw[blue] (3,1)--(3,9)--(4,8);
\draw[blue] (2,13)--(2,12)--(6,8);
\draw[blue] (3,13)--(3,12)--(7,8);
\drawWL {}{4,8}{7,7};
\foreach \x in {5,6,7} {
\drawWL {}{\x,8}{\x-1,8-1};
}

\foreach \x in {1,2,3,4}
{

\node at (4-\x,14) {\x};
\node at (4-\x,0) {\x};
\node at (12-\x,14) {\x};
\node at (12-\x,0) {\x};
}
\end{scope}

\begin{scope}[shift={(5.5,7.5)},xscale=-.5,yscale=-.5]
\draw[purple] (0,3)--(0,13);
\draw[purple] (1,3)--(1,13);
\draw[purple] (2,3)--(2,11)--(5,8);
\draw[purple] (3,3)--(3,9)--(4,8);
\draw[purple] (2,13)--(2,12)--(6,8);
\draw[purple] (3,13)--(3,12)--(7,8);
\end{scope}

\begin{scope}[shift={(2,2.5)},xscale=.5,yscale=.5]
\drawWL {}{4,8}{7,7};
\foreach \x in {5,6,7} {
\drawWL {}{\x,8}{\x-1,8-1};
}
\end{scope}

\begin{scope}[shift={(2,-3)},xscale=.5,yscale=.5]
\drawWL {}{7,8}{4,7};
\foreach \x in {4,5,6} {
\drawWL {}{\x,8}{\x+1,8-1};
}
\end{scope}
\end{scope}
\begin{scope}[shift={(4,2)}]
\node at (0,2) {3D CNOT};
\node at (-3,0) {$\longrightarrow$};
\begin{scope}[shift={(0,0)},xscale=.5,yscale=.5]
\pgftransformcm{1}{0}{0}{1}{\pgfpoint{0}{0}}

\draw[blue!100,dashed] (0,0,-6)--++(0,0,9);
\draw[blue,dashed] (0,0,3)--++(0,0,6);
\draw[blue!100] (0,.8,-6)--++(0,0,9)--+(0,1,0) ;
\draw[blue!100] (1,0,-6)--++(0,0,9);
\draw[blue] (1,0,3)--++(0,0,6);
\draw[blue!100] (1,.8,-6)--++(0,0,10)--+(0,1,0);
\draw[blue] (0,.8,9)--++(0,0,-4)--+(0,1,0);
\draw[blue] (1,.8,9)--++(0,0,-3)--+(0,1,0);
\draw[white,WL] (0,1.8,3)--++(0,1,0)--+(-4,0,-4);
\draw[purple,dashed] (0,1.8,3)--++(0,1,0)--+(-4,0,-4);
\draw[white,WL] (0,1.8,5)--++(0,1,0)--+(-4,0,-4);
\draw[purple] (0,1.8,5)--++(0,1,0)--+(-4,0,-4);
\draw[white,WL] (1,1.8,4)--++(0,1,0)--+(6,0,6);
\draw[purple,dashed] (1,1.8,4)--++(0,1,0)--+(6,0,6);
\draw[white,WL] (1,1.8,6)--++(0,1,0)--+(6,0,6);
\draw[purple] (1,1.8,6)--++(0,1,0)--+(6,0,6);
\draw [white,WL] (-4,3.6,-1)--++(11,0,11);
\draw [purple] (-4,3.6,-1)--++(11,0,11);
\draw [white,WL] (-4,3.6,1)--++(11,0,11);
\draw [purple] (-4,3.6,1)--++(11,0,11);

\node at (0,0,10) {4};
\node at (0,1,10) {1};
\node at (1,0,10) {3};
\node at (1,1,10) {2};

\node at (0,0,-7) {4};
\node at (0,1,-7) {1};
\node at (1,0,-7) {3};
\node at (1,1,-7) {2};

\foreach \x in {6,-6.5}{
\node at (0+\x,1,0+\x) {2};
\node at (0+\x,2,0+\x) {1};
\node at (0+\x,1,2+\x) {3};
\node at (0+\x,2,2+\x) {4};
}

\end{scope}
\end{scope}

\end{scope}
\end{tikzpicture}
\end{center}
%
Earlier 2D representations of approximate multi-qubit CNOT gates  are complicated; some even resemble a musical score, as in Figure 3 of  reference  \cite{Bonesteel-etal}; see also \cite{ZBL}.  Exact  2D qudit CNOT representations appeared in \cite{Hutter-Loss} for odd $d$, where the complexity of the representation depends on $d$.

\section{Resource states}
The generalized Bell states for qudits are given by $B_+=\displaystyle d^{-1/2} \sum_{k\in\Z_{d}}\ket{k,k}$ and  $B_-=\displaystyle d^{-1/2} \sum_{k\in\Z_{d}}\ket{k,-k}$. Diagrammatically:
\vskip -.4cm
\be
\begin{tikzpicture}
\begin{scope}[xscale=.4/1.7,yscale=.2/1.7]
\draw[blue] (0,12) arc (-180:0:2.5);
\draw[blue,dashed] (0,12) arc (180:0:2.5);
\draw[blue] (7,12) arc (-180:0:2.5);
\draw[blue,dashed] (7,12) arc (180:0:2.5);
\draw[blue] (0,12) arc (180:0:6 and 6);
\draw[blue] (5,12) arc (180:0:1 and 1);
\foreach \x in {1,2,3,4}{
\draw (\x,12) arc (180:0:{6-\x});
\node at (\x,11) {\x};
\node at (\x+7,11) {\x};
}
\end{scope}
\end{tikzpicture} \ ,\quad
\begin{tikzpicture}
\begin{scope}[xscale=.4/1.7,yscale=.2/1.7]
\draw[blue] (0,12) arc (-180:0:2.5);
\draw[blue,dashed] (0,12) arc (180:0:2.5);
\draw[blue] (7,12) arc (-180:0:2.5);
\draw[blue,dashed] (7,12) arc (180:0:2.5);
\draw[blue] (0,12) arc (180:0:6 and 6);
\draw[blue] (5,12) arc (180:0:1 and 1);
\foreach \x in {1,2,3,4}{
\draw (\x,12) arc (180:0:{6-\x});
\node at (\x,11) {\x};
}
\node at (8,11) {3};
\node at (9,11) {4};
\node at (10,11) {1};
\node at (11,11) {2};
\end{scope}
\end{tikzpicture} \ .
\ee
The order 3,4,1,2 indicates the action of $F^2$ on the second qudit.
One can check the identifications by the joint relation.
The corresponding multiple-qudit generalizations of the Bell state are known as the  Greenberger-Horne-Zeilinger (GHZ) state \cite{GHZ} and Max. We give their algebraic definitions in \S\ref{Sec:Bi Frobenius algebra}, and their 3D representations in \S \ref{Sect:TopHopf}.

\section{Teleportation}\label{Sect:Teleportation}
In the quon model, we represent the teleportation protocol by the following diagrammatic protocol using the $X$-basis:
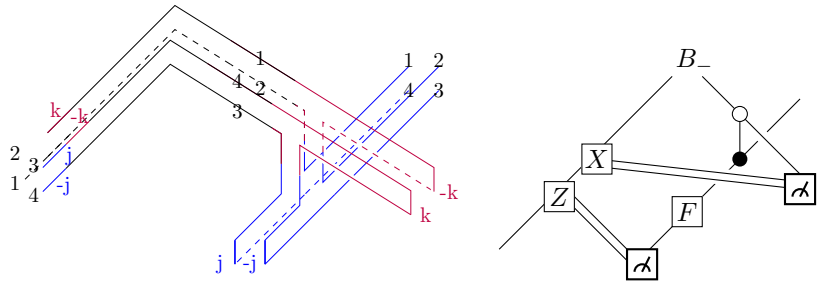
\begin{center}
\raisebox{-.5cm}{
\scalebox{.8}{
\begin{tikzpicture}
\begin{scope}[shift={(17,-6.2)},xscale=.5,yscale=.5]
\pgftransformcm{1}{0}{0}{1}{\pgfpoint{0}{0}}
\draw[blue!100,dashed] (0,0,-6)--++(0,0,9);
\draw[blue,dashed] (0,0,3)--++(0,0,6);
\draw[thick,blue] (0,0,9)--+(0,.8,0);
\node[blue] at (-.5,0,9) {j};
\draw[blue!100] (0,.8,-6)--++(0,0,9)--+(0,1,0) ;
\draw[blue!100] (1,0,-6)--++(0,0,9);
\draw[blue] (1,0,3)--++(0,0,6);
\draw[thick,blue] (1,0,9)--+(0,.8,0);
\node[blue] at (.5,0,9) {-j};
\draw[blue!100] (1,.8,-6)--++(0,0,10)--+(0,1,0);
\draw[blue] (0,.8,9)--++(0,0,-4)--+(0,1,0);
\draw[blue] (1,.8,9)--++(0,0,-3)--+(0,1,0);
\draw[white,WL] (0,1.8,3)--++(0,1,0);
\draw[black,dashed] (0,1.8,3)--++(0,1,0)--+(-4,0,-4);
\draw[purple,dashed] (0,1.8,3)--++(0,1,0)--+(-.5,0,-.5);
\draw[white,WL] (0,1.8,5)--++(0,1,0);
\draw[black] (0,1.8,5)--++(0,1,0)--+(-4,0,-4);
\draw[purple] (0,1.8,5)--++(0,1,0)--+(-.5,0,-.5);
\draw[white,WL] (1,1.8,4)--++(0,1,0)--+(6,0,6);
\draw[purple,dashed] (1,1.8,4)--++(0,1,0)--+(6,0,6);
\draw[white,WL] (1,1.8,6)--++(0,1,0)--+(6,0,6);
\draw[purple] (1,1.8,6)--++(0,1,0)--+(6,0,6);
\draw [white,WL] (7,3.6,10)--++(0,-.8,0);
\draw [purple] (-4,3.6,-1)--++(11,0,11)--++(0,-.8,0);
\draw (-4,3.6,-1)--++(3.5,0,3.5);
\draw [purple] (-4,3.6,1)--++(11,0,11)--++(0,-.8,0);
\draw (-4,3.6,1)--++(3.5,0,3.5);
\node[purple] at (7.5,2.8,12) {k};
\node[purple] at (7.5,2.8,10) {-k};

\draw[dashed] (-4,2.8,-1)--++(-3,0,-3)--++(0,0,11);
\draw (-4,3.6,-1)--++(-3,0,-3)--++(0,0,11);
\draw (-4,2.8,1)--++(-2,0,-2)--++(0,0,7);
\draw (-4,3.6,1)--++(-2,0,-2)--++(0,0,7);

\begin{scope}[shift={(0,0,5)},xscale=1,yscale=1]
\draw[purple] (-7,3.6,0)--++(0,0,2);
\node[purple] at (-7.5,3.6,0) {k};
\draw[purple] (-6,3.6,1)--++(0,0,2);
\node[purple] at (-6.5,3.6,.5) {-k};
\draw (-6,2.8,1)--++(0,0,2);
\draw[dashed] (-7,2.8,0)--++(0,0,2);

\node[blue] at (-5.5,3.6,4) {j};
\draw[blue] (-6,3.6,3)--++(0,0,2);
\node[blue] at (-5.5,2.8,4.5) {-j};
\draw[blue] (-6,2.8,3)--++(0,0,2);
\draw[dashed] (-7,2.8,2)--++(0,0,2);

\node at (0,0,-11) {4};
\node at (0,1,-11) {1};
\node at (1,0,-11) {3};
\node at (1,1,-11) {2};

\node at (-10,0,-3) {1};
\node at (-10,1,-3) {2};
\node at (-9,1,-2) {3};
\node at (-9,0,-2) {4};

\begin{scope}[shift={(2,0,2)}]
\node at (-5,2,-8) {2};
\node at (-5,3,-8) {1};
\node at (-5,3,-6) {4};
\node at (-5,2,-6) {3};
\end{scope}

\end{scope}
\end{scope}
\end{tikzpicture}
}}
\raisebox{-.5cm}{
\scalebox{1}{
\begin{tikzpicture}
\draw (2,-1)--(0,-3);
\draw[white,line width=5pt,opacity=5.0] (.5,-.5)--(2,-2);
\draw (.7,-.7)--(2,-2);
\draw (.3,-.7)--(-2,-3);
\draw (1+.2,-2+.2)--(1+.2,-1-.2);
\fill[black]  (1+.2,-2+.2) circle (.1);
\fill[white]  (1+.2,-1-.2) circle (.1);
\draw[black]  (1+.2,-1-.2) circle (.1);
\fill[white] (.3,-2.3) rectangle (.7,-2.7);
\draw[black] (.3,-2.3) rectangle (.7,-2.7);
\draw (-.25,-3.2)--++(-1,1);
\draw (-.1,-3.2)--++(-1,1);
\draw[white,line width=6pt,opacity=6.0] (2,-2.15)--++(-1.35,.15);
\draw (2,-2.1)--++(-2.7,.3);
\draw (2,-2.2)--++(-2.7,.3);
\node at (.5,-2.5) {$F$};
\begin{scope}[shift={(-1.7,.2)}]
\fill[white] (.3,-2.3) rectangle (.7,-2.7);
\draw[black] (.3,-2.3) rectangle (.7,-2.7);
\node at (.5,-2.5) {$Z$};
\end{scope}
\begin{scope}[shift={(-1.2,.7)}]
\fill[white] (.3,-2.3) rectangle (.7,-2.7);
\draw[black] (.3,-2.3) rectangle (.7,-2.7);
\node at (.5,-2.5) {$X$};
\end{scope}
\begin{scope}[line width=.8pt,shift={(-.1,0)}]
\fill[white] (-.2,-3) rectangle (.2,-3.4);
\draw (-.2,-3) rectangle (.2,-3.4);
\draw (-.1,-3.3) arc (180:0:.1);
\draw (0,-3.3) -- (.1,-3.1);
\end{scope}
\begin{scope}[line width=.8pt,shift={(2,1)}]
\fill[white] (-.2,-3) rectangle (.2,-3.4);
\draw (-.2,-3) rectangle (.2,-3.4);
\draw (-.1,-3.3) arc (180:0:.1);
\draw (0,-3.3) -- (.1,-3.1);
\end{scope}

\node at (.6,-.5) {$B_-$};
\end{tikzpicture}
}}
\end{center}
The pair of oppositely-charged-strings is neutral, thus it is defined in the 3D space. It represents the teleportation process in a topological way.
Moreover, it shows the one-to-one correspondence between the diagrammatic representation of this protocol and the algebraic representation in the teleportation protocol of Bennett et al~\cite{Bennett-etal}, illustrated to the right of the quon-language diagram.

\section{Bi-Frobenius algebras}\label{Sec:Bi Frobenius algebra}
In tensor networks, one decomposes the qudit CNOT gate $C_{X}$ into  COPY and SUM, defined algebraically by planar diagrams.
\beq
CNOT&=&\sum_{k,j\in \Z_d} |k+j,j \rangle \langle k,j |: \quad
\raisebox{-.5cm}{
\begin{tikzpicture}
\begin{scope}[shift={(0,0)},xscale=-.5,yscale=.5]
\draw (0,-1)--(0,1);
\draw (1,-1)--(1,1);
\draw (0,0)--(1,0);
\fill[black] (0,0) circle (.2);
\draw[black] (1,0) circle (.2);
\fill[white] (1,0) circle (.2);
\end{scope}
\end{tikzpicture}}\nonumber\\
COPY&=&\sum_{j\in \Z_d} |j,j \rangle \langle j |:
\raisebox{-.5cm}{
\begin{tikzpicture}
\begin{scope}[shift={(0,0)},xscale=-.5,yscale=.5]
\draw (0,0)--(0,1);
\draw (0,0)--(0,-1);
\draw (0,0)--(1,-1);
\draw[black] (0,0) circle (.2);
\fill[black] (0,0) circle (.2);
\end{scope}
\end{tikzpicture}
}\nonumber\\
SUM&=&\sum_{j\in \Z_d} |k+j \rangle \langle k,j |:
\raisebox{-.5cm}{
\begin{tikzpicture}
\begin{scope}[shift={(0,0)},xscale=-.5,yscale=-.5]
\draw (0,0)--(0,1);
\draw (0,0)--(-1,-1);
\draw (0,0)--(0,-1);
\draw[black] (0,0) circle (.2);
\fill[white] (0,0) circle (.2);
\end{scope}
\end{tikzpicture}
}\nonumber
\eeq
Using the ``spider'' notation of \cite{Coeckebook},
we represent the qudit transformation $\displaystyle \sum_{k\in \Z_d}\overbrace{|k,k,\ldots,k \rangle}^{n \text{ entries}}
\overbrace{ \langle k,k,\ldots, k|}^{n' \text{ entries}}$ by the diagram
\raisebox{-.5cm}{
\begin{tikzpicture}
\begin{scope}[shift={(0,0)},xscale=.5,yscale=.5]
\draw (0,0)--(0,1);
\draw (0,0)--(0,-1);
\draw (0,0)--(-1,-1);
\draw (0,0)--(-1,1);
\draw (0,0)--(1,-1);
\draw (0,0)--(1,1);
\node at (.5,1) {$\cdots$};
\node at (.5,-1) {$\cdots$};
\draw[black] (0,0) circle (.2);
\fill[black] (0,0) circle (.2);
\end{scope}
\end{tikzpicture}
},
and we represent $\displaystyle  \sum_{|k|=|j|} |\vec{k} \rangle \langle \vec{j}|$ by the diagram
\raisebox{-.5cm}{
\begin{tikzpicture}
\begin{scope}[shift={(0,0)},xscale=.5,yscale=-.5]
\draw (0,0)--(0,1);
\draw (0,0)--(0,-1);
\draw (0,0)--(-1,-1);
\draw (0,0)--(-1,1);
\draw (0,0)--(1,-1);
\draw (0,0)--(1,1);
\node at (.5,1) {$\cdots$};
\node at (.5,-1) {$\cdots$};
\draw[black] (0,0) circle (.2);
\fill[white] (0,0) circle (.2);
\end{scope}
\end{tikzpicture}
}.
In particular, one represents the state $\ket{0}$ by
\raisebox{-.3cm}{
\begin{tikzpicture}
\begin{scope}[shift={(0,0)},xscale=.5,yscale=.5]
\draw (0,0)--(0,-2);
\draw[black] (0,0) circle (.2);
\fill[white] (0,0) circle (.2);
\end{scope}
\end{tikzpicture}}\;.
We have the duality induced by the Fourier transform $F$ between the two  spiders.
\vskip -.5cm
\be \label{Equ:duality}
\raisebox{-1.2cm}{
\scalebox{.7}{
\begin{tikzpicture}
\begin{scope}[xscale=1.5,yscale=1.5]
\begin{scope}[shift={(0,0)},xscale=.5,yscale=.5]
\node at (.5,-2) {$\cdots$};
\draw (0,0)--(0,-1)--++(0,-1);
\draw (0,0)--(-1,-1)--++(0,-1);
\draw (0,0)--(1,-1)--++(0,-1);
\draw[black] (0,0) circle (.2);
\fill[white] (0,0) circle (.2);
\end{scope}
\begin{scope}[shift={(0,0)},xscale=.5,yscale=-.5]
\node at (.5,-2) {$\cdots$};
\draw (0,0)--(0,-1)--++(0,-1);
\draw (0,0)--(-1,-1)--++(0,-1);
\draw (0,0)--(1,-1)--++(0,-1);
\draw[black] (0,0) circle (.2);
\fill[white] (0,0) circle (.2);
\end{scope}
\end{scope}
\end{tikzpicture}}}
=d^{n/2-1}
\raisebox{-1.2cm}{
\scalebox{.8}{
\begin{tikzpicture}
\begin{scope}[xscale=1.5,yscale=1.5]
\begin{scope}[shift={(2.5,0)},xscale=.5,yscale=.5]
\node at (.5,-2) {$\cdots$};
\draw (0,0)--(0,-1)--++(0,-1);
\draw (0,0)--(-1,-1)--++(0,-1);
\draw (0,0)--(1,-1)--++(0,-1);
\draw[black] (0,0) circle (.2);
\fill[black] (0,0) circle (.2);

\draw[black] (0,-1) circle (.4);
\fill[white] (0,-1) circle (.4);
\node at (0,-1) {$F$};

\draw[black] (-1,-1) circle (.4);
\fill[white] (-1,-1) circle (.4);
\node at (-1,-1) {$F$};

\draw[black] (1,-1) circle (.4);
\fill[white] (1,-1) circle (.4);
\node at (1,-1) {$F$};
\end{scope}
\begin{scope}[shift={(2.5,0)},xscale=.5,yscale=-.5]
\node at (.5,-2) {$\cdots$};
\draw (0,0)--(0,-1)--++(0,-1);
\draw (0,0)--(-1,-1)--++(0,-1);
\draw (0,0)--(1,-1)--++(0,-1);
\draw[black] (0,0) circle (.2);
\fill[black] (0,0) circle (.2);

\draw[black] (0,-1) circle (.4);
\fill[white] (0,-1) circle (.4);
\node at (0,-1) {$F^{-1}$};

\draw[black] (-1,-1) circle (.4);
\fill[white] (-1,-1) circle (.4);
\node at (-1,-1) {$F^{-1}$};

\draw[black] (1,-1) circle (.4);
\fill[white] (1,-1) circle (.4);
\node at (1,-1) {$F^{-1}$};
\end{scope}
\end{scope}
\end{tikzpicture}}}
\ ,
\ee
where $n$ is the number of boundary points. This generalizes the duality between the resource states analyzed in \cite{JLW-HS}:
\[
\Max=d^{\frac{1-n}{2}} \sum_{|\vec{k}|=0} | \vec{k} \rangle\;,\quad
\GHZ=\frac{1}{\sqrt{d}} \sum_{k\in \Z_d}| k,k, \cdots ,k \rangle\;.
\]
 In particular, one represents $\displaystyle \sum_{k=0}^{d-1} \ket{k}=\sqrt{d}$
\raisebox{-.5cm}{
\begin{tikzpicture}
\begin{scope}[shift={(0,0)},xscale=.5,yscale=.5]
\draw (0,0)--(0,-.5);
\draw (-.5,-.5) rectangle (.5,-1.5);
\node at (0,-1) {$F$};
\draw (0,-1.5)--(0,-2);
\draw[black] (0,0) circle (.2);
\fill[white] (0,0) circle (.2);
\end{scope}
\end{tikzpicture}} \
by
\raisebox{-.5cm}{
\begin{tikzpicture}
\begin{scope}[shift={(0,0)},xscale=.5,yscale=.5]
\draw (0,0)--(0,-2);
\fill[black] (0,0) circle (.2);
\end{scope}
\end{tikzpicture}}\ .

The adjoint transformations are represented by the vertical reflections of these diagrams.
It is known that both trivalent vertices are Frobenius algebras. That means the following relations hold for
\raisebox{-.5cm}{
\scalebox{.8}{
\begin{tikzpicture}
\begin{scope}[shift={(0,0)},xscale=.5,yscale=.5]
\draw (0,0)--(0,-1);
\draw (0,-1)--(-1,-2);
\draw (0,-1)--(1,-2);
\fill[black] (0,-1) circle (.2);
\end{scope}
\end{tikzpicture}}},
\raisebox{-.5cm}{
\scalebox{.8}{
\begin{tikzpicture}
\begin{scope}[shift={(0,0)},xscale=.5,yscale=.5]
\draw (0,0)--(0,-1);
\draw[black] (0,0) circle (.2);
\fill[black] (0,0) circle (.2);
\end{scope}
\end{tikzpicture}}}\;,
and similarly for
\raisebox{-.5cm}{
\scalebox{.8}{
\begin{tikzpicture}
\begin{scope}[shift={(0,0)},xscale=.5,yscale=.5]
\draw (0,0)--(0,-1);
\draw (0,-1)--(-1,-2);
\draw (0,-1)--(1,-2);
\draw[black] (0,-1) circle (.2);
\fill[white] (0,-1) circle (.2);
\end{scope}
\end{tikzpicture}}},
\raisebox{-.5cm}{
\scalebox{.8}{
\begin{tikzpicture}
\begin{scope}[shift={(0,0)},xscale=.5,yscale=.5]
\draw (0,0)--(0,-1);
\draw[black] (0,0) circle (.2);
\fill[white] (0,0) circle (.2);
\end{scope}
\end{tikzpicture}}}\;:

\be
\raisebox{-.5cm}{
\scalebox{.8}{
\begin{tikzpicture}
\begin{scope}[shift={(0,0)},xscale=.5,yscale=.5]
\draw (0,0)--(0,-1);
\draw (0,-1)--(-2,-3);
\draw (0,-1)--(2,-3);
\draw (-1,-2)--(0,-3);
\fill[black] (0,-1) circle (.2);
\fill[black] (-1,-2) circle (.2);
\end{scope}
\end{tikzpicture}}}
=
\raisebox{-.5cm}{
\scalebox{.8}{
\begin{tikzpicture}
\begin{scope}[shift={(0,0)},xscale=.5,yscale=.5]
\draw (0,0)--(0,-1);
\draw (0,-1)--(-2,-3);
\draw (0,-1)--(2,-3);
\draw (1,-2)--(0,-3);
\fill[black] (0,-1) circle (.2);
\fill[black] (1,-2) circle (.2);
\end{scope}
\end{tikzpicture}}}\;,
\ee

\be
\raisebox{-.5cm}{
\scalebox{.8}{
\begin{tikzpicture}
\begin{scope}[shift={(0,0)},xscale=-.5,yscale=.5]
\draw (0,0)--(0,1);
\draw (0,0)--(0,-1);
\draw (0,0)--(1,-1);
\fill[black] (1,-1) circle (.2);
\fill[black] (0,0) circle (.2);
\end{scope}
\end{tikzpicture}}}
\ =
\raisebox{-.5cm}{
\scalebox{.9}{
\begin{tikzpicture}
\begin{scope}[shift={(0,0)},xscale=-.5,yscale=.5]
\draw (0,-1)--(0,1);
\end{scope}
\end{tikzpicture}}}\;.
\ee
One can flip the boundary points of a black/white spider from top to bottom or the other way using caps or cups labeled by a black/white bullet.
Thus each Frobenius algebra has a compatible pivotal structure. However, the two Frobenius algebras do not share pivotal structures.
A composition of the cap and the cup with different colored bullets is not the identity map. Instead it is the antipode map $\FS^2=F^2$, where $\FS$ is the string Fourier transform defined in \S\ref{Sec:TopAlg},
\be \label{Equ:antipode}
\raisebox{-.25cm}{
\scalebox{.8}{
\begin{tikzpicture}
\begin{scope}[shift={(0,0)},xscale=-.5,yscale=.5]
\draw (0,0)--(1,1)--(2,0)--(3,1);
\fill[black] (1,1) circle (.2);
\fill[white] (2,0) circle (.2);
\draw[black] (2,0) circle (.2);
\end{scope}
\end{tikzpicture}}}
=
\raisebox{-.25cm}{
\scalebox{.8}{
\begin{tikzpicture}
\begin{scope}[shift={(0,0)},xscale=-.5,yscale=.5]
\draw (0,0)--(1,1)--(2,0)--(3,1);
\fill[black] (2,0) circle (.2);
\fill[white] (1,1) circle (.2);
\draw[black] (1,1) circle (.2);
\end{scope}
\end{tikzpicture}}}
=
\raisebox{-.25cm}{
\scalebox{.8}{
\begin{tikzpicture}
\begin{scope}[shift={(0,0)},xscale=-.5,yscale=.5]
\draw (0,-1)--(0,1);
\fill[white] (0,0) circle (.5);
\draw (0,0) circle (.5);
\node at (0,0) {$F^2$};
\end{scope}
\end{tikzpicture}}}\;.
\ee

In the quon model, we can represent these maps in a consistent way by strings in 3-manifolds, such that the algebraic relations become topological isotopy.
The COPY map is represented by
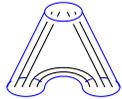
\[
\begin{tikzpicture}
\begin{scope}[shift={(4,-3)},xscale=.5,yscale=.5]
\draw[blue] (-.5,-1) to [bend left=80] (.5,-1);
\draw[blue] (-1.5,-1) arc (-180:0:.5 and .25);
\draw[blue,dashed] (-1.5,-1) arc (180:0:.5 and .25);
\draw[blue] (1.5,-1) arc (0:-180:.5 and .25);
\draw[blue,dashed] (1.5,-1) arc (0:180:.5 and .25);
\draw (-1.3,-1)--(-.3,1);
\draw (-1.1,-1)--(-.1,1);
\draw (1.3,-1)--(.3,1);
\draw (1.1,-1)--(.1,1);
\draw (-.7,-1) to [bend left=80] (.7,-1);
\draw (-.9,-1) to [bend left=80] (.9,-1);
\draw[white,WL] (.5,1) arc (0:360:.5 and .25);
\draw[blue] (.5,1) arc (0:360:.5 and .25);
\draw[blue] (-1.5,-1)--(-.5,1);
\draw[blue] (1.5,-1)--(.5,1);
\end{scope}
\end{tikzpicture}.
\]
One can check that this diagrammatic definition coincides with the algebraic definition by \eqref{Equ:joint}.
By this representation, the algebraic conditions of the Frobenius algebra become topological isotopy in the quon model.

\section{The String-Genus Joint Relation} \label{Sect:String-Genus}
In case we have a diagram of the form: a closed neutral string surrounds a genus of the manifold, then we can remove both, up to a scalar.
When $m$ and $n$ are odd numbers, we have
\be \label{Equ:circle-genus}
\scalebox{.9}{$
\raisebox{-1.4cm}{
\scalebox{.65}{
\begin{tikzpicture}
\begin{scope}[shift={(0,0)},xscale=1.5,yscale=.75]
\draw[blue] (0,0) arc (-180:0:1 and .5);
\draw[blue,dashed] (0,0) arc (180:0:1 and .5);
\draw[blue] (0,4) arc (-180:0:1 and .5);
\draw[blue,dashed] (0,4) arc (180:0:1 and .5);
\draw[blue] (0.5,2) arc (-180:0:.5 and .25);
\draw[blue] (.7,1.8) arc (180:0:.3 and .25);
\draw[blue] (0,0) to [bend right=-30] (0,4);
\draw[blue] (2,0) to [bend right=30] (2,4);
\draw[thick] (0.3,0) to [bend right=-30] (0.3,4);
\draw[thick] (1.7,0) to [bend right=30] (1.7,4);
\draw[thick] (0.6,0) to [bend right=-30] (0.6,4);
\draw[thick] (1.4,0) to [bend right=30] (1.4,4);
\node at (-.1,2) {$\cdots$};
\node at (2.1,2) {$\cdots$};
\draw (.3,2) arc (180:-180:.7 and .5);
\node at (.4,-.6) {$m$};
\node at (1.6,-.6) {$n$};
\end{scope}
\end{tikzpicture}}}
=d^{-1/2}
\raisebox{-1.4cm}{
\scalebox{.65}{
\begin{tikzpicture}
\begin{scope}[shift={(0,0)},xscale=1.5,yscale=.75]
\draw[blue] (0,0) arc (-180:0:1 and .5);
\draw[blue,dashed] (0,0) arc (180:0:1 and .5);
\draw[blue] (0,4) arc (-180:0:1 and .5);
\draw[blue,dashed] (0,4) arc (180:0:1 and .5);
\draw[blue] (0,0) to [bend right=-30] (0,4);
\draw[blue] (2,0) to [bend right=30] (2,4);
\draw[thick] (0.3,0) to [bend right=-30] (0.3,4);
\draw[thick] (1.7,0) to [bend right=30] (1.7,4);
\draw[thick] (0.6,0) to [bend right=-30] (0.6,4);
\draw[thick] (1.4,0) to [bend right=30] (1.4,4);
\node at (-.1,2) {$\cdots$};
\node at (2.1,2) {$\cdots$};
\node at (.4,-.6) {$m$};
\node at (1.6,-.6) {$n$};
\end{scope}
\end{tikzpicture}}}$}
\ee
Note that $\tau^{m}$, $\tau^{n}$ are multiples of $\tau$. It is enough to prove the relation for $m=n=1$. In this case, it follows from the relation \eqref{Equ:joint}.
%
%
%

If $m$ or $n$ is even,  by relation \eqref{Equ:joint}, then the diagram is 0. Thus if the diagram is a part of a non-zero transformation, then the  $m$ and $n$ have to be odd numbers.
In this case, the relation means that if there is a circle around a genus of the 3-manifold, then we can remove the circle and the genus by multiplying a scalar $d^{-1/2}$.

\section{Topological Relations for ${C^{*}}$-Hopf Algebras}\label{Sect:TopHopf}
A more conceptual way to look at pictures in the quon model is to assign  the four boundary points of the quon to the corners of a square in a plane orthogonal to the $Z$-axis.
The white and black bullets in the spiders indicate diagrammatic operations in $X$ and $Y$-directions on the 2D plane.
For example, if we look at this 3D diagram from the top along the $Z$-direction, then the picture for the $Z$-basis in \S\ref{Sect:1-QuonBasis} is given by
\[
\frac{1}{\sqrt{d}}
\scalebox{.6}{
\raisebox{-1cm}{
\begin{tikzpicture}
\foreach \x in {0,1}{
\foreach \y in {0,1}{
\coordinate (A\x\y) at (\x,\y);
\node at (A\x\y) {$\bullet$};
\draw[white,WL] (.5,.5) circle (1);
\draw[blue!50] (.5,.5) circle (1);
}}
\draw (0,1) arc (-180:0:.5 and .3);
\draw (0,0) arc (180:0:.5 and .3);
\node at (.5,1) {-k};
\node at (.5,0) {k};
\end{tikzpicture}}}\ ,
\]
for $k\in \Z_d$.
Here we only draw the boundary circle of the 3-manifold to simplify the picture.
The pictures for COPY and SUM become respectively
\begin{align*}
\raisebox{-1.5cm}{
\scalebox{.9}{
\begin{tikzpicture}
\begin{scope}[xscale=.65,yscale=.65]
\foreach \x in {0,1}{
\foreach \y in {0,1}{
\foreach \u in {-1}{
\foreach \v in {-1,1}{
\coordinate (A\u\v\x\y) at (\x+1.5*\u,\y+1.5*\v);
\node at (A\u\v\x\y) {$\bullet$};
\draw[white,WL] (1.5*\u+.5,1.5*\v+.5) circle (1);
\draw[blue!50] (1.5*\u+.5,1.5*\v+.5) circle (1);
}}}}
\foreach \x in {0,1}{
\foreach \y in {0,1}{
\foreach \u in {-1}{
\foreach \v in {0}{
\coordinate (A\u\v\x\y) at (.5*\x+2*\u+.75,.5*\y+2*\v+.25);
}}
}}
\draw[white,WL] (A-1001)--(A-1101);
\draw[white,WL] (A-1011)--(A-1111);
\draw[white,WL] (A-1000)--(A-1-100);
\draw[white,WL] (A-1010)--(A-1-110);
\draw[white,WL] (A-1-101) to [bend  left=30] (A-1100);
\draw[white,WL] (A-1-111) to [bend  right=30] (A-1110);
\draw[purple,1WL] (A-1001)--(A-1101);
\draw[purple,1WL] (A-1011)--(A-1111);
\draw[purple,1WL] (A-1000)--(A-1-100);
\draw[purple,1WL] (A-1010)--(A-1-110);
\draw[purple,1WL] (A-1-101) to [bend  left=30] (A-1100);
\draw[purple,1WL] (A-1-111) to [bend right=30] (A-1110);
\foreach \x in {0,1}{
\foreach \y in {0,1}{
\foreach \u in {-1}{
\foreach \v in {0}{
\node at (A\u\v\x\y) {$\bullet$};
\draw[white,WL] (2*\u+1,2*\v+.5) circle (.6);
\draw[blue!100] (2*\u+1,2*\v+.5) circle (.6);
}}}}
\end{scope}
\end{tikzpicture}}}
& \text{~and~} d^{-1/2}
\raisebox{-1.3cm}{
\scalebox{.9}{
\begin{tikzpicture}
\begin{scope}[xscale=.45,yscale=.45]
\foreach \x in {0,1}{
\foreach \y in {0,1}{
\foreach \u in {-1,1}{
\foreach \v in {-1}{
\coordinate (A\u\v\x\y) at (\x+1.5*\u,\y+1.5*\v);
\node at (A\u\v\x\y) {$\bullet$};
}}}}
\foreach \x in {0,1}{
\foreach \y in {0,1}{
\foreach \u in {0}{
\foreach \v in {-1}{
\coordinate (A\u\v\x\y) at (3*\x+2*\u-1,3*\y+2*\v-.5);
\node at (A\u\v\x\y) {$\bullet$};
\draw[blue!30] (2*\u+.5,2*\v+1) circle (3);
}}
}}
\draw[white,WL] (A0-101)--(A-1-101);
\draw[white,WL] (A0-100)--(A-1-100);
\draw[white,WL] (A0-111)--(A1-111);
\draw[white,WL] (A0-110)--(A1-110);
\draw[white,WL] (A-1-111)--(A1-101);
\draw[white,WL] (A-1-110)--(A1-100);
\draw[red,1WL] (A0-101)--(A-1-101);
\draw[red,1WL] (A0-100)--(A-1-100);
\draw[red,1WL] (A0-111)--(A1-111);
\draw[red,1WL] (A0-110)--(A1-110);
\draw[red,1WL] (A-1-111)--(A1-101);
\draw[red,1WL] (A-1-110)--(A1-100);
\foreach \x in {0,1}{
\foreach \y in {0,1}{
\foreach \u in {-1,1}{
\foreach \v in {-1}{
\node at (A\u\v\x\y) {$\bullet$};
\draw[white,WL] (1.5*\u+.5,1.5*\v+.5) circle (1);
\draw[blue!100] (1.5*\u+.5,1.5*\v+.5) circle (1);
}}}}
\end{scope}
\end{tikzpicture}}}\ .
\end{align*}

Similarly we can represent the white and black spiders as strings in 3-manifolds in the quon model. The white spiders are expended in the $X$-direction and the black spiders are expended in the $Y$-direction.
In particular, the resource states $\GHZ$ and $\Max$ (for 3-quons) are respectively:
\begin{align*}
\raisebox{-1.5cm}{
\begin{tikzpicture}
\begin{scope}[xscale=.5,yscale=.5]
\foreach \x in {0,1}{
\foreach \y in {0,1}{
\foreach \u in {0}{
\foreach \v in {-2,0,2}{
\coordinate (A\u\v\x\y) at (\x+1.2*\u,\y+1.2*\v);
\node at (A\u\v\x\y) {$\bullet$};
\draw[blue!50] (1.2*\u+.5,1.2*\v+.5) circle (1);
}}}}
\draw (A0-200) to [bend right=12] (A0201);
\draw (A0-210) to [bend left=12] (A0211);
\draw (A0-201) -- (A0000);
\draw (A0001) -- (A0200);
\draw (A0-211) -- (A0010);
\draw (A0011) -- (A0210);
\end{scope}
\end{tikzpicture}
}
\quad , \quad
\raisebox{-.5cm}{
\begin{tikzpicture}
\begin{scope}[xscale=.5,yscale=.5]
\foreach \x in {0,1}{
\foreach \y in {0,1}{
\foreach \u in {-2,0,2}{
\foreach \v in {0}{
\coordinate (A\v\u\y\x) at (\x+1.2*\u,\y+1.2*\v);
\node at (A\v\u\y\x) {$\bullet$};
\draw[blue!50] (1.2*\u+.5,1.2*\v+.5) circle (1);
}}}}
\draw (A0-200) to [bend right=-12] (A0201);
\draw (A0-210) to [bend left=-12] (A0211);
\draw (A0-201) -- (A0000);
\draw (A0001) -- (A0200);
\draw (A0-211) -- (A0010);
\draw (A0011) -- (A0210);
\end{scope}
\end{tikzpicture}
}.
\end{align*}

From this point of view, the Fourier transform is a $90^\circ$ rotation around the $Z$-axis, which explains the duality of the two Frobenius algebras in \eqref{Equ:duality} in a geometric way.
Moreover, one can check that the relation \eqref{Equ:antipode} also becomes an isotopy in the 3D space in the quon model.

Furthermore, this pair of Frobenius algebras satisfies the following additional relations \eqref{Equ:hopf 2}--(\ref{Equ:hopf}). These relations define $\Z_d$ as a $C^*$ Hopf algebra, where $F^2$ is the antipode map of the Hopf algebra and the involution is an anti-linear map which reflects the diagrams vertically. Note that if a pair of Frobenius algebras satisfy these relations, then the underlying $d$ dimensional Hilbert space becomes a Hopf algebra. This has been observed in \cite{Lafont,Coeckebook}.

\begin{align}
\label{Equ:hopf 2}
\scalebox{.7}{
\raisebox{-.5cm}{
\begin{tikzpicture}
\begin{scope}[shift={(0,0)},xscale=.5,yscale=.5]
\draw (0,-2)--(0.5,-1)--(0.5,0);
\draw (1,-2)--(0.5,-1);
\fill[black] (.5,-1) circle (.2);
\draw[black] (.5,0) circle (.2);
\fill[white] (.5,0) circle (.2);
\end{scope}
\end{tikzpicture}}}
&=
\scalebox{.7}{
\raisebox{-.5cm}{
\begin{tikzpicture}
\begin{scope}[shift={(0,0)},xscale=.5,yscale=.5]
\draw (0,0)--(0,-2);
\draw (1,0)--(1,-2);
\draw[black] (0,0) circle (.2);
\fill[white] (0,0) circle (.2);
\draw[black] (1,0) circle (.2);
\fill[white] (1,0) circle (.2);
\end{scope}
\end{tikzpicture}}} \\
\label{Equ:hopf 3}
\scalebox{.7}{
\raisebox{-.5cm}{
\begin{tikzpicture}
\begin{scope}[shift={(0,0)},xscale=.5,yscale=-.5]
\draw (0,-2)--(0.5,-1)--(0.5,0);
\draw (1,-2)--(0.5,-1);
\fill[black] (.5,0) circle (.2);
\draw[black] (.5,-1) circle (.2);
\fill[white] (.5,-1) circle (.2);
\end{scope}
\end{tikzpicture}}}
&=
\scalebox{.7}{
\raisebox{-.5cm}{
\begin{tikzpicture}
\begin{scope}[shift={(0,0)},xscale=.5,yscale=-.5]
\draw (0,0)--(0,-2);
\draw (1,0)--(1,-2);
\fill[black] (0,0) circle (.2);
\fill[black] (1,0) circle (.2);
\end{scope}
\end{tikzpicture}}}
\\
\label{Equ:hopf 1}
\scalebox{.7}{
\raisebox{-1cm}{
\begin{tikzpicture}
\begin{scope}[shift={(0,0)},xscale=.5,yscale=.5]
\draw (0.5,-3)--(0.5,-2);
\draw (0.5,0)--(0.5,1);
\draw (0.5,0) to [bend left=60] (0.5,-2);
\draw (0.5,0) to [bend left=-60] (0.5,-2);
\fill[black] (0.5,0) circle (.2);
\draw[black] (0.5,-2) circle (.2);
\fill[white] (0.5,-2) circle (.2);
\fill[white] (0,-1) circle (.5);
\draw (0,-1) circle (.5);
\node at (0,-1) {$F^2$};
\end{scope}
\end{tikzpicture}}}
&=
\scalebox{.7}{
\raisebox{-1cm}{
\begin{tikzpicture}
\begin{scope}[shift={(0,0)},xscale=.5,yscale=.5]
\draw (0.5,-3)--(0.5,-2);
\draw (0.5,0)--(0.5,1);
\draw (0.5,0) to [bend left=60] (0.5,-2);
\draw (0.5,0) to [bend left=-60] (0.5,-2);
\fill[black] (0.5,0) circle (.2);
\draw[black] (0.5,-2) circle (.2);
\fill[white] (0.5,-2) circle (.2);
\fill[white] (1,-1) circle (.5);
\draw (1,-1) circle (.5);
\node at (1,-1) {$F^2$};
\end{scope}
\end{tikzpicture}}}
=
\scalebox{.7}{
\raisebox{-.6cm}{
\begin{tikzpicture}
\begin{scope}[shift={(0,0)},xscale=.5,yscale=.5]
\draw (0.5,-2)--(0.5,-1);
\draw (0.5,0)--(0.5,1);
\fill[black] (0.5,0) circle (.2);
\draw[black] (0.5,-1) circle (.2);
\fill[white] (0.5,-1) circle (.2);
\end{scope}
\end{tikzpicture}}}\\\label{Equ:hopf}
\scalebox{.7}{
\raisebox{-.6cm}{
\begin{tikzpicture}
\begin{scope}[shift={(0,0)},xscale=.5,yscale=.5]
\draw (0,-2)--(0,1);
\draw (1,0)--(0,-1);
\draw (1,-2)--(1,1);
\draw (0,0)--(1,-1);
\fill[black] (0,0) circle (.2);
\fill[black] (1,0) circle (.2);
\draw[black] (0,-1) circle (.2);
\fill[white] (0,-1) circle (.2);
\draw[black] (1,-1) circle (.2);
\fill[white] (1,-1) circle (.2);
\end{scope}
\end{tikzpicture}}}
\ &=
\scalebox{.7}{
\raisebox{-.6cm}{
\begin{tikzpicture}
\begin{scope}[shift={(0,0)},xscale=.5,yscale=.5]
\draw (0,-2)--(0.5,-1)--(0.5,0)--(0,1);
\draw (1,-2)--(0.5,-1)--(0.5,0)--(1,1);
\fill[black] (.5,-1) circle (.2);
\draw[black] (.5,0) circle (.2);
\fill[white] (.5,0) circle (.2);
\end{scope}
\end{tikzpicture}}}
\end{align}
%

\bigskip\noindent

Now we give an interpretation of these relations as topological isotopy in 3D.  Note that relation \eqref{Equ:hopf 2} follows from the definition of the COPY map.
By isotopy in our quon model, relations \eqref{Equ:hopf 3} and \eqref{Equ:hopf 1} are exactly the same diagrams as the relation \eqref{Equ:hopf 2}.  The most interesting relation is \eqref{Equ:hopf}, and we explain that relation in detail.

Relation \eqref{Equ:hopf} becomes topological isotopy, when we use the string-genus joint relation established in \S\ref{Sect:String-Genus},  namely the joint relation \eqref{Equ:circle-genus}.  Then \eqref{Equ:hopf} is given by the isotopy \eqref{Hopf-Relation-identity}. Thus we have given a topological interpretation for the $C^{*}$ Hopf algebra axioms for~$\Z_{d}$.
\begin{align}\label{Hopf-Relation-identity}
\raisebox{-1.4cm}{
\scalebox{.8}{
\begin{tikzpicture}
\draw[blue!40] (.25,2.2) arc (90:-270: 2 and .7);
\draw[blue!40] (.25,-.05) arc (90:-270: 2 and .7);
\draw[blue] (.05,0.4) arc (180:0:.2 and .2);
\draw[blue] (-.05,0.5) arc (-180:0:.3 and .3);
\begin{scope}[xscale=.75,yscale=.75]
\foreach \x in {0,1}{
\foreach \y in {0,1}{
\foreach \u in {-1}{
\foreach \v in {-1,1}{
\coordinate (A\u\v\x\y) at (\x+1.5*\u,\y+1.5*\v);
\node at (A\u\v\x\y) {$\bullet$};
}}}}
\draw[red] (A-1-110)--++(1.6,0);
\draw[red] (A-1-111)--++(1.6,0);
\draw[red] (A-1110)--++(1.6,0);
\draw[red] (A-1111)--++(1.6,0);
\foreach \x in {0,1}{
\foreach \y in {0,1}{
\foreach \u in {-1}{
\foreach \v in {0}{
\coordinate (A\u\v\x\y) at (.5*\x+2*\u+.75,.5*\y+2*\v+.25);
}}
}}
\draw[white,WL] (A-1001)--(A-1101);
\draw[white,WL] (A-1011)--(A-1111);
\draw[white,WL] (A-1000)--(A-1-100);
\draw[white,WL] (A-1010)--(A-1-110);
\draw[white,WL] (A-1-101)--(A-1100);
\draw[white,WL] (A-1-111)--(A-1110);
\draw[purple] (A-1001)--(A-1101);
\draw[purple] (A-1011)--(A-1111);
\draw[purple] (A-1000)--(A-1-100);
\draw[purple] (A-1010)--(A-1-110);
\draw[purple] (A-1-101) to [bend  left=30] (A-1100);
\draw[purple] (A-1-111) to [bend right=30] (A-1110);
\foreach \x in {0,1}{
\foreach \y in {0,1}{
\foreach \u in {-1}{
\foreach \v in {0}{
\node at (A\u\v\x\y) {$\bullet$};
\draw[white,WL] (2*\u+1,2*\v+.5) circle (.6);
\draw[blue!80] (2*\u+1,2*\v+.5) circle (.6);
}}}}
\end{scope}
\begin{scope}[shift={(2,0)},xscale=.75,yscale=.75]
\foreach \x in {0,1}{
\foreach \y in {0,1}{
\foreach \u in {-1}{
\foreach \v in {-1,1}{
\coordinate (A\u\v\x\y) at (\x+1.5*\u,\y+1.5*\v);
\node at (A\u\v\x\y) {$\bullet$};
}}}}
\foreach \x in {0,1}{
\foreach \y in {0,1}{
\foreach \u in {-1}{
\foreach \v in {0}{
\coordinate (A\u\v\x\y) at (.5*\x+2*\u+.75,.5*\y+2*\v+.25);
}}
}}
\draw[white,WL] (A-1001)--(A-1101);
\draw[white,WL] (A-1011)--(A-1111);
\draw[white,WL] (A-1000)--(A-1-100);
\draw[white,WL] (A-1010)--(A-1-110);
\draw[white,WL] (A-1-101)--(A-1100);
\draw[white,WL] (A-1-111)--(A-1110);
\draw[purple] (A-1001)--(A-1101);
\draw[purple] (A-1011)--(A-1111);
\draw[purple] (A-1000)--(A-1-100);
\draw[purple] (A-1010)--(A-1-110);
\draw[purple] (A-1-101) to [bend  left=30] (A-1100);
\draw[purple] (A-1-111) to [bend right=30] (A-1110);
\foreach \x in {0,1}{
\foreach \y in {0,1}{
\foreach \u in {-1}{
\foreach \v in {0}{
\node at (A\u\v\x\y) {$\bullet$};
\draw[white,WL] (2*\u+1,2*\v+.5) circle (.6);
\draw[blue!80] (2*\u+1,2*\v+.5) circle (.6);
}}}}
\end{scope}
\end{tikzpicture}}}
=d^{-1/2}&
\raisebox{-1.4cm}{
\scalebox{.8}{
\begin{tikzpicture}
\draw[red] (-.5,0.575)--++(1,0);
\draw[red] (-.5,0.175)--++(1,0);
\draw[blue!40] (0,2.2) arc (90:-270: 1.5 and .7);
\draw[blue!40] (0,-.05) arc (90:-270: 1.5 and .7);
\begin{scope}[xscale=.75,yscale=.75]
\foreach \x in {0}{
\foreach \y in {0,1}{
\foreach \u in {-1}{
\foreach \v in {-1,1}{
\coordinate (A\u\v\x\y) at (\x+1.5*\u,\y+1.5*\v);
\node at (A\u\v\x\y) {$\bullet$};
}}}}
\foreach \x in {0,1}{
\foreach \y in {0,1}{
\foreach \u in {-1}{
\foreach \v in {0}{
\coordinate (A\u\v\x\y) at (.5*\x+2*\u+.75,.5*\y+2*\v+.25);
}}
}}
\draw[white,WL] (A-1001)--(A-1101);
\draw[white,WL] (A-1000)--(A-1-100);
\draw[white,WL] (A-1-101)--(A-1100);
\draw[purple] (A-1001)--(A-1101);
\draw[purple] (A-1000)--(A-1-100);
\draw[purple] (A-1-101) to [bend  left=30] (A-1100);
%
\foreach \x in {0,1}{
\foreach \y in {0,1}{
\foreach \u in {-1}{
\foreach \v in {0}{
\node at (A\u\v\x\y) {$\bullet$};
\draw[white,WL] (2*\u+1,2*\v+.5) circle (.6);
\draw[blue!80] (2*\u+1,2*\v+.5) circle (.6);
}}}}
\end{scope}
\begin{scope}[shift={(1.5,0)},xscale=.75,yscale=.75]
\foreach \x in {1}{
\foreach \y in {0,1}{
\foreach \u in {-1}{
\foreach \v in {-1,1}{
\coordinate (A\u\v\x\y) at (\x+1.5*\u,\y+1.5*\v);
\node at (A\u\v\x\y) {$\bullet$};
}}}}
\foreach \x in {0,1}{
\foreach \y in {0,1}{
\foreach \u in {-1}{
\foreach \v in {0}{
\coordinate (A\u\v\x\y) at (.5*\x+2*\u+.75,.5*\y+2*\v+.25);
}}
}}
\draw[white,WL] (A-1011)--(A-1111);
\draw[white,WL] (A-1010)--(A-1-110);
\draw[white,WL] (A-1-111)--(A-1110);
\draw[purple] (A-1011)--(A-1111);
\draw[purple] (A-1010)--(A-1-110);
\draw[purple] (A-1-111) to [bend right=30] (A-1110);
\foreach \x in {0,1}{
\foreach \y in {0,1}{
\foreach \u in {-1}{
\foreach \v in {0}{
\node at (A\u\v\x\y) {$\bullet$};
\draw[white,WL] (2*\u+1,2*\v+.5) circle (.6);
\draw[blue!80] (2*\u+1,2*\v+.5) circle (.6);
}}}}
\end{scope}
\end{tikzpicture}}}
\end{align}

There is a one-to-one correspondence between $C^{*}$ Hopf algebras and irreducible, depth-two subfactor planar algebras \cite{Szy94,KLS03}.
In this case, $d$ is the global dimension of the $C^{*}$-Hopf algebra. Moreover, the even part of the planar algebra is the representation category of the Kac algebra. The odd part has only one simple object $\tau$, so \eqref{Equ:circle-genus} also holds.
In the above interpretation, we only use (shaded) planar diagrams without braids in 3-manifolds. So this topological interpretation works for any finite dimensional $C^{*}$ Hopf algebra.
From this point of view, many algebraic properties of $C^{*}$-Hopf algebras reduce to topological isotopy.

\section{Quon Language for  a unitary modular tensor category}\label{Sect:UMTC}
We can define the quon language for any unitary modular tensor category $\mathscr{C}$, so that the 1-quon basis corresponds to the set of simple objects $OB$ in $\mathscr{C}$.
If we take $\mathscr{C}$ to be the unitary modular tensor category, such that its fusion ring is $Z_d$ and its modular $S$ matrix is $q^{kl}$, where $q=e^{\frac{2\pi i}{d}}$, then we get back the quon language for qudits defined by PAPPA.

For each $X\in OB$, we obtain a simple object $\tilde{X}:=X\otimes \overline{X}$ in $\mathscr{C}\otimes \mathscr{C}$, where $\overline{X}$ is the dual of $X$.
Take $\gamma=\bigoplus_{X\in OB} X\otimes \overline{X}$ in $\mathscr{C}\otimes \mathscr{C}$.
It is known that $\gamma=\bigoplus_{X\in OB} X\otimes \overline{X}$ is a Frobenius algebra in $\mathscr{C}\otimes \mathscr{C}$.
Thus $\mathscr{P}_{n,+}=\hom(1,\gamma^n)$ is a subfactor planar algebra generated by $\tau$, such that $\gamma=\tau\otimes \overline{\tau}$. It is proved in \cite{LX} that this planar algebra is unshaded. That means $\tau=\overline{\tau}$.

Note that
$\hom(1,\tau^4)\cong \hom(1,\gamma^2)$. This space has an orthonormal-basis given by the canonical inclusion from $1$ to $(X\otimes \overline{X}) \otimes (\overline{X} \otimes X)$, for $X\in OB$.
We call the generalized single-particle state a  1-quon.
Thus the dimension $d$ of the 1-quon space is the cardinality of $OB$.
Furthermore,  the string Fourier transform on the 1-quon space is the $S$ matrix of the unitary modular tensor category $\mathscr{C}$ \cite{LX}.

We generalize the quon language as follows: We label each of the four boundary points by the object $\tau$. The diagrams in the 3-manifolds are given by morphisms in $\mathscr{C}\otimes \mathscr{C}$.
The representation for an $n$-quon is given by morphisms in $\hom(1,\tau^{4})$ in $n$ hemispheres as in \eqref{Equ:qudit}.
The $n$-quon transformations are represented by morphisms in $\hom (\tau^{4n},\tau^{4n})$ in a 3-manifold as in \eqref{Equ:T}.
The relations between diagrams and 3-manifolds are also defined by  \eqref{Equ:joint}.
Then $n$-quon transformations also reduce to a linear sum of the form in \eqref{Equ:transformation-basis} which represent matrix units. Therefore the $n$-quon transformations are transformations on the $d^n$ dimensional Hilbert space.

In general, the quon language can be defined for any subfactor planar algebra \cite{Jones}, if we do not require $\tau=\overline{\tau}$. In this case, the diagrams in the 3-manifolds with $4n$ boundary points are given by a shaded planar diagram in the $2n$-box space of planar algebras. We have used this general case to give the topological interpretation of the $C^{*}$-Hopf algebra relations.

\section{Acknowledgment}
We thank Daniel Loss for discussions about~\cite{Hutter-Loss} and Michael Freedman and Kevin Walker for discussion about~\cite{Freedman-etal-2}. We thank Bob Coecke for sharing the book manuscript \cite{Coeckebook}.   We thank all three referees for helpful, constructive comments.  A.J. and Z.L.  thank the Max Planck Institute for Mathematics, Bonn and the Hausdorff Research Institute for Mathematics, Bonn for hospitality where during the initiation of this work.
This research was supported by a grant from the Templeton Religion Trust.


\begin{thebibliography}{1}
\bibitem{Kitaev03}
A.\ Kitaev,
Fault-tolerant quantum computation by anyons,
\textit{Ann. Phys.} {\bf 303} (2003), 2--30,
\blue{\url{https://arxiv.org/abs/quant-ph/9707021}},
{\color{blue} \url{10.1016/S0003-4916(02)00018-0}}.

\bibitem{Freedman-etal}
M.\ H.\ Freedman, A.\ Kitaev, M.\ J.\ Larsen, and Z.\ Wang,
Topological quantum computation,
\textit{Bulletin of the American Mathematical Society} Volume 40, Number 1, (2002), 31--38,
{\color{blue} \url{http://dx.doi.org/10.1090/S0273-0979-02-00964-3}}.

\bibitem{JL}
A.\ M.\ Jaffe and Z.\ Liu,
Planar para algebras, reflection positivity, {\it Commun. Math. Phys.} to appear (2017),
\blue{\url{http://arxiv.org/abs/1602.02662}}.

\bibitem{JLW-QI}
A.\ M.\ Jaffe, Z.\ Liu, and A.\ Wozniakowski,
Qudit isotopy,
\blue{\url{https://arxiv.org/abs/1602.02671}}.


\bibitem{JLW-HS}
A.\ M.\ Jaffe, Z.\ Liu, and A.\ Wozniakowski,
Holographic software for quantum networks, preprint,
\blue{\url{http://arxiv.org/abs/1605.00127}}.

\bibitem{JLW-TDP}
A.\ M.\ Jaffe, Z.\ Liu, and A.\ Wozniakowski,
Constructive simulation and topological design of protocols, 
\textit{New Journal of Physics},
in print  (2017),
\blue{\url{http://iopscience.iop.org/article/10.1088/1367-2630/aa5b57/pdf}}

\bibitem{Kitaev}
A.\ Kitaev,
Anyons in an exactly solved model and beyond,
\textit{Ann. Phys.} {\bf 306} (2006) 2--111,
\blue{\url{http://dx.doi.org/10.1016/j.aop.2005.10.005}}.


\bibitem{Penrose}
R. Penrose,  
Application of negative dimension tensors,  
pp. 221--244, in
``Combinatorial Mathematics and its Applications,''
Edited by D.J.A.~Welsh, Academic Press (1971).


\bibitem{Deutsch}
D.~Deutsch, 
Quantum computational networks.
\textit{Proceedings of the Royal Society of London. Series A, Mathematical and Physical Sciences},
Vol. 425, No. 1868  (1989), 73--90.

\bibitem{DVC}
W.~D\"ur, G.~Vidal, and J.~I.~ Cirac,  
Three qubits can be entangled in two inequivalent ways.
\textit{Phys. Rev. A} {\bf 62},  (2000) 062314.

\bibitem{Lafont}
Y.~Lafont,  
Towards an algebraic theory of Boolean circuits.
\textit{Jour. Pure Appl. Algebra}, {\bf 184}, (2003) 257--310.


\bibitem{AbramskyCoecke04}
S.\ Abramsky and B.\ Coecke, A categorical semantics of quantum protocols,
\textit{Logic in Computer Science, 2004. Proceedings of the 19th Annual IEEE Symposium} IEEE (2004).



\bibitem{Coecke-Duncan-1}
B.~Coecke and R.~Duncan,
Interacting quantum observables. In \textit{Automata, Languages and Programming}, {\bf 5126} 298--310 (2008),  Springer,
\blue{\url{http://dx.doi.org/10.1007/978-3-540-70583-3 25}}.

\bibitem{Coecke-Duncan-2}
B.~Coecke and R.~Duncan,
Interacting quantum observables: categorical algebra and diagrammatics,
\textit{New Journal of Physics}, {\bf 13(4)}  043016 (2011),
\blue{\url{http://dx.doi.org/10.1088/1367-2630/13/4/043016}}.

\bibitem{Coeckebook}
B.\ Coecke and A.\ Kissinger,
Picturing quantum processes:
A first course in quantum theory and diagrammatic reasoning,
Cambridge University Press, Cambridge, UK, to appear (2017),
{ISBN:9781107104228}.

\bibitem{BB}
V. Bergholm and J. Biamonte,
Categorical quantum circuits,
\textit{J. Phys A: Math. Theor.} {\bf 44} 245304 (2011),
\blue{\url{https://arxiv.org/abs/1010.4840}}.


\bibitem{DBJC}
S.~J.~Denny J.~D.~Biamonte, D.~Jaksch, and S.~R.~Clark  Algebraically contractible topological tensor network states, \textit{J. Phys. A: Math. Theor.}  {\bf 45} 015309 (2012),
\blue{\url{https://arxiv.org/abs/1108.0888}}.


\bibitem{Backens}
M.~K.~Backens,
Completeness and the $ZX$-calculus,
Thesis, Oxford University,
\blue{\url{https://arxiv.org/abs/1602.08954}}.

\bibitem{Vicary}
J.~Vicary,
Higher quantum theory,
\blue{\url{https://arxiv.org/abs/1207.4563}}.

\bibitem{Vicary-Reutter}
D.~Reutter and J.~Vicary,
Biunitary constructions in quantum information,
\blue{\url{https://arxiv.org/abs/1609.07775}}.

\bibitem{Jones}
V.~F.~R.~Jones,
Planar algebras I.,
(1999),
\blue{\url{https://arxiv.org/abs/math/9909027}}.

\bibitem{KL-1}
L.\ Kauffman and S.\ Lomonaco Jr.,
Quantum entanglement and topological entanglement,
\textit{New J. Phys.} {\bf 4}, (2002) 73.1--73.18, 
{\color{blue} \doi{10.1088/1367-2630/4/1/373}}.

\bibitem{RHG}
R. Raussendorf, J.~Harrington, and K.~Goyal,
A fault-tolerant one-way quantum computer,
\textit{Annals of Physics} {\bf 321}, (2006) 2242--2270,
{\color{blue} \url{http://www.dx.doi.org/10.1016/j.aop.2006.01.012}}

\bibitem{HFDV}
C.~Horsman, A.~G.~Fowler, S.~Devitt, and R.~Van~Meter,
Surface code quantum computing by lattice surgery,
\textit{New Jour. Phys.}
{\bf 14}, (2012) 123011,
 {\color{blue} \url{http://www.dx.doi.org/10.1088/1367-2630/14/12/123011}}

\bibitem{BLKW}
B.~J.~Brown, K.~Laubscher, M/~S.~Kesselring, and J.~R.~Wootton,
Poking holes and cutting corners to achieve Clifford gates with the surface code,
{\color{blue}\url{https://arxiv.org/abs/1609.04673}}.

\bibitem{Bonesteel-etal}
N.~E.~Bonesteel, L.~Hormozi, G.~Zikos, and S.~H.~Simon,
Braid Topologies for Quantum Computation,
\textit{Phys. Rev. Lett.} {\bf 95}, 140503 (2005),
{\color{blue}\url{http://dx.doi.org/10.1103/PhysRevLett.95.140503}}.

\bibitem{ZBL}
O.~Zilberberg, B.~Braunecker, and D.~ Loss,
Controlled-NOT gate for multiparticle qubits and topological quantum computation based on parity measurements,
\textit{Phys. Rev. A} {\bf 77}, 012327 (2008),
{\color{blue}\url{http://dx.doi.org/10.1103/PhysRevA.77.012327}}.

\bibitem{Hutter-Loss}
A.\ Hutter and D.\ Loss,
Quantum computing with parafermions,
\textit{Phys. Rev.} {\bf B 93}, (2016) 125105 1--7,
\blue{\url{http://dx.doi.org/10.1103/PhysRevB.93.125105}}.

\bibitem{GHZ}
D.\ M.\ Greenberger, M.\ A.\ Horne, and A.\ Zeilinger,
Going beyond Bell's theorem,
in \textit{Bell's theorem, quantum theory, and conceptions of the universe}, M.\ Kafakos, editor, Vol. 37 of ``Fundamental Theories of Physics,'' Springer Verlag, Heidelberg (1989),
{\color{blue}\url{http://arxiv.org/abs/0712.0921}}
{\color{blue}\url{http://link.springer.com/chapter/10.1007/978-94-017-0849-4_10}}.


\bibitem{Bennett-etal}
C.\ H.\ Bennett, G.\ Brassard, C.\ Cr\'epeau, R.\ Jozsa, A.\ Peres, and W.\ K.\ Wootters,
Teleporting an unknown quantum state via dual classical and Einstein-Podolsky-Rosen Channels, \textit{Phys. Rev. Lett.} {\bf 70}, (1993)   1895,
{\color{blue} \url{http://dx.doi.org/10.1103/PhysRevLett.70.1895}}.

\bibitem{Szy94}
W.~Szymanski,
Finite index subfactors and Hopf algebra crossed products
\textit{Proc. Amer. Math. Soc.} {\bf 120}, (1994) no. 2, 519-528.


\bibitem{KLS03}
V.~Kodiyalam, Z.~Landau, and V.~S.~Sunder,
The planar algebra associated to a Kac algebra,
\textit{Proceedings of the Indian Academy of Sciences-Mathematical Sciences.} (2003) Vol. 113. No. 1. Springer India.

\bibitem{LX}
Z. Liu, and F. Xu,
Jones-Wassermann subfactors for modular tensor categories,
{\color{blue}\url{https://arxiv.org/abs/1612.08573}}.


\bibitem{Freedman-etal-2}
T.~Karzig, C.~ Knapp, R.~M.~Lutchyn, P~Bonderson, M.~Hastings, C.~Nayak, J.~Alicea, K.~Flensberg, S.~Plugge, Y.~Oreg, C.~Marcus, and M.~H.~Freedman,
Scalable designs for quasiparticle-poisoning-protected topological quantum computation with Majorana zero modes,
{\color{blue}\url{https://arxiv.org/abs/1610.05289}}.

\end{thebibliography}
\end{document}